\documentclass[showpacs,showkeys,preprint,amsmath,amssymb,nofootinbib, prd, aps, 11pt]{revtex4-1}
\usepackage[T1]{fontenc}
\usepackage[latin1]{inputenc}

\usepackage[sc, osf]{mathpazo}
\usepackage[mathscr]{eucal}
\usepackage{framed, color}
\definecolor{shadecolor}{rgb}{0.84,0.84,0.84}
\usepackage{graphicx}
\newcommand{\vek}[1]{\mathbf{#1}}

\newcommand{\ab}{\mathsf{a}}
\newcommand{\db}{\mathsf{d}}

\newcommand{\dd}{\text{\dj}}

\newcommand{\da}{\mathsf{d}}

\begin{document}


\title{Covariant variational approach to Yang-Mills Theory: Thermodynamics}

\author{M.~Quandt}\email{markus.quandt@uni-tuebingen.de}
\author{H.~Reinhardt}\email{hugo.reinhardt@uni-tuebingen.de}
\affiliation{%
Universit\"at T\"ubingen\\
Institut f\"ur Theoretische Physik\\
Auf der Morgenstelle 14\\
D-72076 T\"ubingen, Germany
}%

\date{\today}


\begin{abstract}
\noindent
The thermodynamics of $SU(2)$ Yang-Mills theory in the covariant 
variational approach is studied by relating the free action density 
in the background of a non-trivial Polyakov loop to the pressure of 
the gluon plasma. The correct subtraction of the vacuum contribution 
in the free action density is argued for. The Poisson resummed 
expression for the pressure can be evaluated analytically in limiting cases, 
and shows the correct Stefan-Boltzmann limit at $T \to \infty$, while 
the limit $T \to 0$ is afflicted by artifacts due to massless 
modes in the confined phase. Several remedies to remove these artifacts are 
discussed. Using the numerical $T=0$ solutions for the ghost and gluon 
propagators in the covariant variational approach as input, the pressure, 
energy density and interaction strength are calculated and compared to 
lattice data.
\end{abstract}


\maketitle


\section{Introduction}
\label{sec:intro}
The low energy sector of quantum chromodynamics (QCD) and, in particular, 
its phase diagram are among the most actively researched topics in 
elementary particle physics. While heavy ion collisions at the large hadron 
collider (LHC) now begin to explore in detail the quark-gluon plasma 
at large temperatures and net baryon densities, the theoretical description of the 
full phase diagram through lattice simulations is still hampered by the sign 
problem. Alternative continuum methods are therefore of particular 
interest.

Among the most powerful continuum techniques are the so-called \emph{functional
methods} which, in one way or the other, amount to a closed set of integral
equations connecting the low-order Green's functions of the theory. 
For instance, the Dyson-Schwinger equations (\emph{DSE}) are merely the 
equations of motion for the Green's functions in momentum space, truncated 
by arbitrary (or educated) assumptions about higher vertices
\cite{Fischer:2006ub,Alkofer:2000wg,Binosi:2009qm}. Alternatively, 
one- or two-loop quantum corrections about non-perturbative extensions of the 
usual Faddeev-Popov (\emph{FP}) action by mass terms 
\cite{Tissier:2010ts,*Tissier:2011ey} or the Gribov-Zwanziger term 
\cite{Zwanziger:1989mf, *Zwanziger:1992qr} have also been discussed 
\cite{Canfora:2015yia}. A more elaborated technique 
are the functional renormalization group (\emph{FRG}) flow equations 
\cite{Pawlowski:2005xe,Gies:2006wv} which describe the behaviour of the relevant 
operators in the effective action as the theory evolves from the cutoff scale 
towards its infrared fixed point. Finally, if we are willing to dispense 
with manifest covariance, a particularly appealing and physically transparent 
picture emerges in the Hamiltonian approach to QCD in Coulomb gauge using 
variational techniques \cite{Feuchter:2004gb,Reinhardt:2004mm,Epple:2006hv}.

Recently, we have added an alternative continuum approach to the list
\cite{Quandt:2013wna,Quandt:2015aaa,Quandt:2016ykm} 
which attempts to combine the insightfulness of the Hamiltonian approach with the 
simplicity of a manifestly covariant setup. The method is based on  
\emph{Ans\"atze} for the euclidean path integral measure, and results in a 
closed set of integral equations that can be conventionally renormalized. 
The numerical solutions give excellent agreement with zero-temperature lattice 
propagators \cite{Quandt:2013wna}, and also offer a satisfying description of 
the finite-temperature corrections, which agree with expectations from the 
lattice in all qualitative aspects \cite{Quandt:2015aaa}. Recently, the method
has also been used to study the deconfinement phase transition where it predicts 
the correct order of the transition for the colour groups $SU(2)$ and $SU(3)$ and 
gives quantitative values for the transition temperature which are in fair
agreement with lattice data \cite{Quandt:2016ykm}. 

In the present paper, we want to complete the study of the Yang-Mills (\emph{YM})
thermodynamics by computing its equation of state. The covariant variational
approach is particularly well suited for this investigation, since it gives 
direct access to the free action density (and hence the pressure) of the 
gluon plasma. This is much more complicated in some of the functional methods 
mentioned above which only give (partial) information on the \emph{derivatives}
of the effective action. Nontheless, the equation of state has been 
computed in a variety of functional methods before: in the DSE approach, 
for instance, the phase structure of full QCD including quarks has been 
explored, both at finite temperature and finite density 
\cite{Fischer:2011mz, Fischer:2012vc}.
The pressure in pure Yang-Mills theory has also been calculated using massive 
extensions of the Faddeev-Popov action \cite{Reinosa:2014zta, Reinosa:2015gxn}, 
where comparison with lattice data gives a decent agreement in the deconfined phase, 
but deviations in the confined phase. In an alternative approach, the 
poor convergence of the perturbation series in the deconfined phase 
can be stabilized by non-perturbative resummation techniques \cite{Fukushima:2013xsa}, 
which yield a good description of the pressure in this region, but lack e.g.~the peak 
structure in the interaction strength.
Finally, the pressure has also been computed using continuum methods in Coulomb 
gauge. A first approach using the Gribov formula as the temperature-independent 
dispersion relation for a gas of free gluon excitations gave results comparable
to massive free particles while lacking a true phase transition \cite{Zwanziger:2004np}. 
In Ref.~\cite{Reinhardt:RX6}, good agreement with lattice data could be achieved
within the variational Hamiltonian approach, 
provided that the grand canonical density operator was constructed from a full 
ensemble of approximate thermal states above the vacuum. 
In an alternative Hamiltonian formulation, the thermodynamics can also be related to 
the ground state properties of YM-theory on a semi-compactified space manifold 
$\mathbb{R}^2 \times S^1$ \cite{Reinhardt:2016xci}. In this case, the pressure exhibits 
significant deviations from the expected behaviour \cite{Reinhardt:RX8}. This can 
be traced to the violation of $O(4)$ invariance underlying the spatial 
compactification.

This paper is organised as follows: In the next section, we review the 
covariant variation principle and show how the free energy can naturally
be accessed within this approach. Section \ref{sec:var} summarizes the previous
results obtained for YM theory and, in particular, the computational technique 
for the Polyakov loop study \cite{Quandt:2016ykm}, which will also be used 
(with appropriate modifications) for the present investigation. The necessary 
background field technique is described in more detail in section \ref{sec:background},
where we also discuss the correct subtraction of the vacuum energy and 
the analytical limits of our solution. Section \ref{sec:results} presents our 
numerical results for the various thermodynamical quantities in comparison
to lattice data. In the last sub-section \ref{sec:discuss}, we discusses the 
origin of deviations and artifacts in the confined region, and speculate 
about possible remedies.  We conclude the manuscript with a brief
summary and an outlook on future investigations within the present approach.

\section{Thermodynamics and the variation principle }
\label{sec:thermo}
Let us briefly recall the variation principle for the effective action
of pure Yang-Mills theory in Euclidean spacetime. The variation
is with respect to the normalized path integral measure $ d\mu(A)$ which is used 
to compute expectation values of arbitrary observables built from the gauge field $A \equiv A_\mu^a(x)$. 
Within the space of such probability measures, quantum field theory singles out the 
particular Gibb's type of measure
\begin{align}
d\mu_0[A] = Z^{-1}\cdot \mathcal{D} A\, \mathcal{J}[A]\, 
\exp\Big\{- \hbar^{-1}\,S_{\rm fix}[A]\Big\}\,,
\label{gibbs}
\end{align}
where $S_{\rm fix}=S_{\rm YM} + S_{\rm gf}$ is the classical euclidean action including the 
gauge-fixing term, and $\mathcal{J}[A]$ is the corresponding Faddeev-Popov (FP) determinant. 
(We do not need to specify the gauge condition at this point.) The partition function
\begin{align}
 Z \equiv\int \mathcal{D}A\,\mathcal{J}[A]\, \exp\Big\{- \hbar^{-1}\,S_{\rm fix}[A]\Big\}
\end{align}
is required for normalization. Although it is formally a gauge-dependent quantity, 
the Faddeev-Popov procedure, or Becchi-Rouet-Stora-Tyutin (BRST) symmetry, entails 
that $Z$ agrees in all local gauges. The moments of the Gibbs-measure $d\mu_0$ from 
eq.~(\ref{gibbs})  are the usual Schwinger functions of euclidean field theory. 
Moreover, Gibb's measure minimizes the \emph{free action}
\begin{align}
F(\mu) \equiv \big\langle\,S_{\rm fix}\,\big\rangle_\mu - \hbar \,
\mathcal{W}(\mu)
\label{freeaction}
\end{align}
where the relative entropy
\begin{align}
\mathcal{W}(\mu) = - \left\langle\,\ln(\rho / \mathcal{J})\,\right\rangle_\mu\,.
\label{relentrop}
\end{align}
describes the available phase space for quantum fluctuations in the trial measure
$d\mu \equiv \mathcal{D}A\,\rho[A]$
relative to the FP determinant, which is the natural measure on the gauge orbit.
Moreover, the absolute minimum of the free action is given by 
\begin{align}
\min_{\mu}\,F(\mu) = F(\mu_0) = -\ln Z\,,
\label{minF}
\end{align}
which is hence also a gauge-invariant quantity.
It is convenient to perform the minimization of
the free action in two steps, by first constraining $F(\mu)$ such that the 
expectation value of an arbitrary operator $\Omega[A]$ is fixed at a prescribed 
classical value $\omega$. The minimum is then called the 
\emph{effective action} for the operator $\Omega$,
\begin{align}
\Gamma[\omega] = \min_\mu\,\Big\{ F(\mu) \,\big\vert\,\langle\,\Omega \,\rangle_\mu = \omega \Big\}\,.
\label{var1}
\end{align}
The most common choice is to take $\Omega[A]$ as the quantum gauge field itself
(with classical value $\langle A\rangle_\mu = \mathscr{A}$)
whence the derivatives of $\Gamma[\mathscr{A}]$ become the proper functions 
of the full quantum  theory. The FP procedure will only ensure that 
$\Gamma[\omega]$ is gauge invariant, \emph{if} $\Omega$ happens to be a 
gauge-invariant operator; in particular, the effective action 
$\Gamma[\mathscr{A}]$ is a gauge-dependent functional, and so are the 
proper functions. However, the absolute \emph{minimum} of 
$\Gamma[\mathscr{A}]$ is related to the free energy by 
\begin{align}
\min_{\mathscr{A}}\,\Gamma[\mathscr{A}] = F(\mu_0) = - \ln Z
\label{Gammamin} 
\end{align}
and must hence agree in all gauges, provided that no approximations have been 
made.

\medskip\noindent
To study the thermodynamics of the system, we can introduce finite temperature 
using the imaginary time formalism, i.e.~we compactify the Euclidean time direction 
to the finite interval $[0, \beta]$ and impose periodic boundary conditions in this 
direction for both the gluons and the ghosts \cite{Bernard:1974bq}. This will make all 
quantities introduced above depend on the inverse temperature $\beta$. In particular,
the minimum of the free action can now be written
\begin{align}
\min_\mu F_\beta(\mu)=\min_{\mathscr{A}} \Gamma_\beta[\mathscr{A}] 
= - \ln Z(\beta) \equiv V_3 \,\beta\cdot f(\beta, V_3)\,,
\end{align}
where $V_3$ is the spatial volume.
In gauges which do not break the residual spatial $O(3)$ symmetry explicitly, 
the density $f$ depends on $V_3$ only through the spatial boundary conditions, 
which are expected to become immaterial in the infinite volume limit, i.e.~the
quantity
\[
f(\beta) \equiv \lim_{V_3 \to \infty} f(\beta, V_3) 
\]
should exist for such gauges and, by gauge invariance of $Z$, eventually 
also in all other local gauges. Other thermodynamical quantities can be deduced 
directly from $f(\beta)$: in a grand canonical ensemble, the chemical potential 
for the (massless) gluons must vanish and we have the relations listed in the 
following table: 
\medskip
\renewcommand{\arraystretch}{1.2} 
\begin{table}[h!]
\centering
\begin{tabular}{c|l}
\toprule
pressure & $\quad p(\beta) = - f(\beta)$ 
\\
energy density & $\quad \epsilon(\beta) = f + \beta \,\partial f/ \partial \beta$
\\
entropy density & $ \quad w(\beta) = \beta(\epsilon + p) = \beta^2 \,\partial f / \partial \beta$ 
\\
interaction strength \quad & $ \quad \Delta(\beta) = \beta^4(\epsilon - 3 p) = -\beta\,\partial 
(p \beta^4) / \partial \beta$
\\ \botrule
\end{tabular}
\label{tab:1}
\end{table}
\renewcommand{\arraystretch}{1.0}

The interaction strength measures the deviation from the ideal gas limit, since 
$p \sim T^4$ and hence $\Delta = 0$ for an ultra-relativistic gas at vanishing chemical 
potential. The only independent thermodynamical information about the gluon 
plasma is thus contained in the \emph{equation of state}, i.e.~the pressure 
function $p(\beta)$, which is the main target of the present investigation.

\section{The variational approach}
\label{sec:var}
So far, all considerations apply to the exact minimum of the free action 
eq.~(\ref{freeaction}) attained for the Gibbs measure eq.~(\ref{gibbs}), 
or, equivalently, for the exact minimum of the effective action (\ref{var1}). 
The \emph{variational method} restricts the space of trial measures $d\mu$ 
to some subset of normalized measures for which (i) the constraint 
$\langle \Omega\rangle_\mu = \omega$ can be implemented conveniently, and 
(ii) the relevant expectation values can be evaluated reliably. 

As a first step, we have to fix a gauge. At zero temperature, the unbroken 
spacetime $O(4)$ invariance entails that the most natural choice is 
\emph{Landau gauge}. This is not because the effective action 
$\Gamma[\mathscr{A}]$ would be particularly easy to calculate in this gauge --- 
it is, in fact, a very hard calculation for classical fields 
$\mathscr{A}_\mu^a$ that have arbitrary colour and Lorentz orientation
\cite{Quandt:2013wna}. The real reason for using Landau gauge is that the 
manifestly unbroken spacetime $O(4)$ invariance entails $\langle A_\mu \rangle = 0$,
i.e.~the effective action \emph{must} take its minimum at the classical field 
$\mathscr{A}_\mu = 0$. This allows us to concentrate the variational search
eq.~(\ref{var1}) to measures with vanishing first moment 
$\langle A\rangle_\mu = 0$, which simplifies the calculation considerably.

\begin{figure}[t]
\centering
\includegraphics[width=6cm,clip]{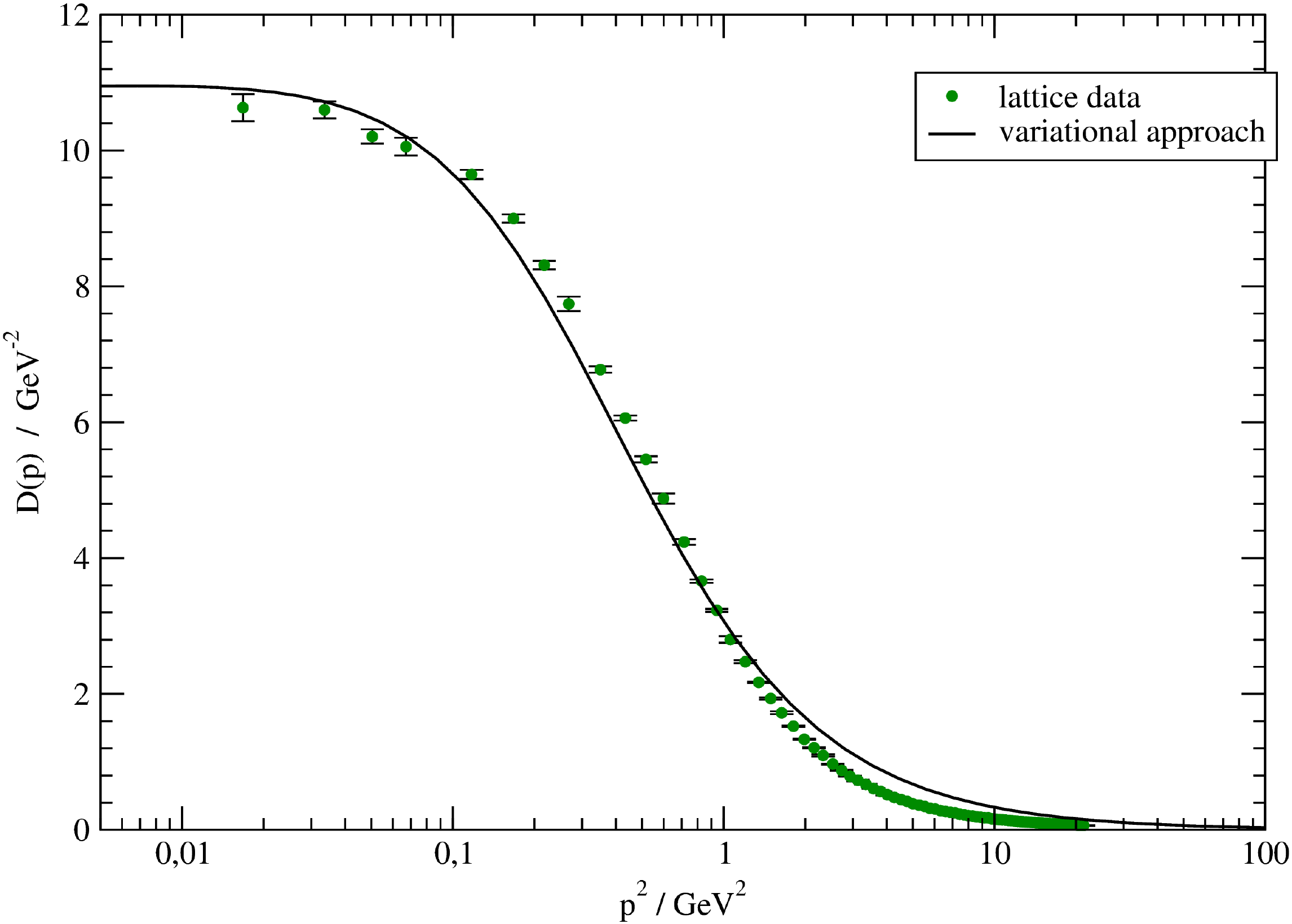}
\hspace*{1cm}
\includegraphics[width=6cm,clip]{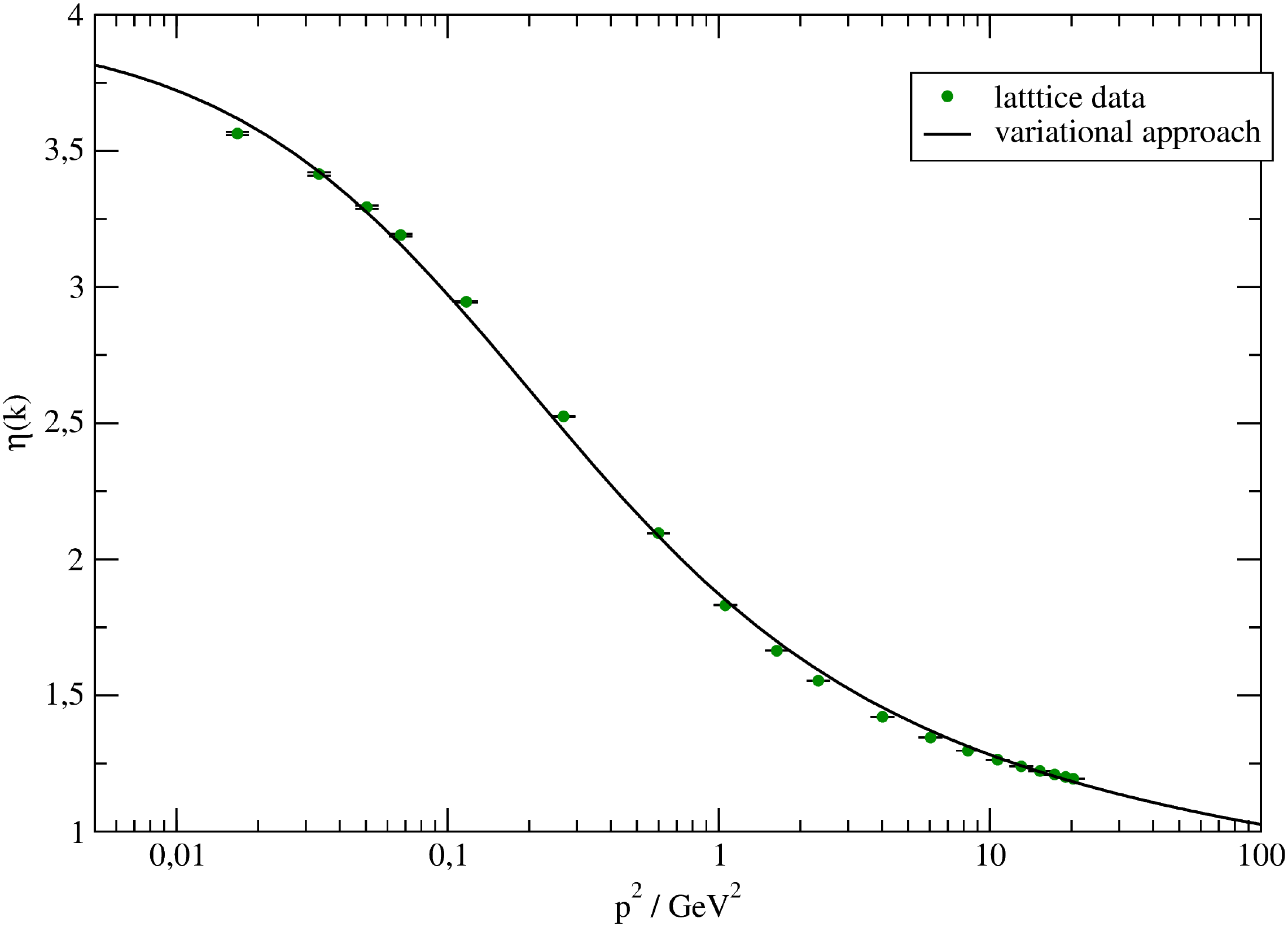}
\caption{The zero-temperature gluon propagator (\emph{left}) and the ghost 
form factor (\emph{right}) for the decoupling type of solution, 
compared to high-precision lattice data \cite{Bogolubsky:2009dc}.}
\label{fig:1}       
\end{figure}

In Refs.~\cite{Quandt:2013wna, Quandt:2015aaa}, we have thus made a 
simple yet physically sensible ansatz and considered only Gaussian measures 
of the form
\begin{align}
d\mu[A] = \mathcal{N} \cdot \mathcal{D}A\,\mathcal{J}[A]^{1-2\alpha}\cdot 
\exp\Bigg\{ - \frac{1}{2}\int d(x,y)\, A_\mu^a(x)\,\omega_{\mu\nu}^{ab}(x,y)\,A^b_\nu(y)\Bigg\}\,,
\label{gauss}
\end{align}
where the constant $\alpha$ and the kernel $\omega$ are variational parameters.
Physically, the picture conveyed by this ansatz is that of a weakly interacting 
(constituent) gluon with a enhanced weight (for $\alpha > 0$) near the Gribov horizon. 
If the FP determinant 
\begin{align}
\mathcal{J}[A] = \frac{\mathrm{det}\big(- \partial_\mu \hat{D}_\mu\big)}{\mathrm{det}(-\partial^2)}\,. 
\end{align}
is treated to the same formal loop order as the remaining exponent in eq.~(\ref{gauss}), it 
can be replaced by the simpler expression
\begin{align}
\ln \mathcal{J}[A] \approx - \frac{1}{2}\,\int d(x,y)\,A_\mu^a(x)\,\chi^{ab}_{\mu\nu}(x,y)
\,A_\nu^b(y)\,,
\label{curv}
\end{align}
where the \emph{curvature} $\chi$ can be expressed through the Faddeev-Popov ghost operator 
$G = (\partial_\mu \hat{D}^\mu)^{-1}$ and the bare ($\Gamma_0$) and full ($\Gamma$) ghost-gluon 
vertex, 
\begin{align}
\chi^{ab}_{\mu\nu}(x,y) = - \left\langle \frac{\delta^2 \ln \mathcal{J}}{\delta A_\mu^a(x)\,
\delta A_\nu^b(y)}\right \rangle = - \mathrm{Tr}\Big[\, \langle G \rangle \,\Gamma_\mu^a(x)\,
\langle G \rangle\, \Gamma_{0,\nu}^{b}(y)\,\Big] = 
\quad \begin{minipage}{3cm}\includegraphics[width=3cm]{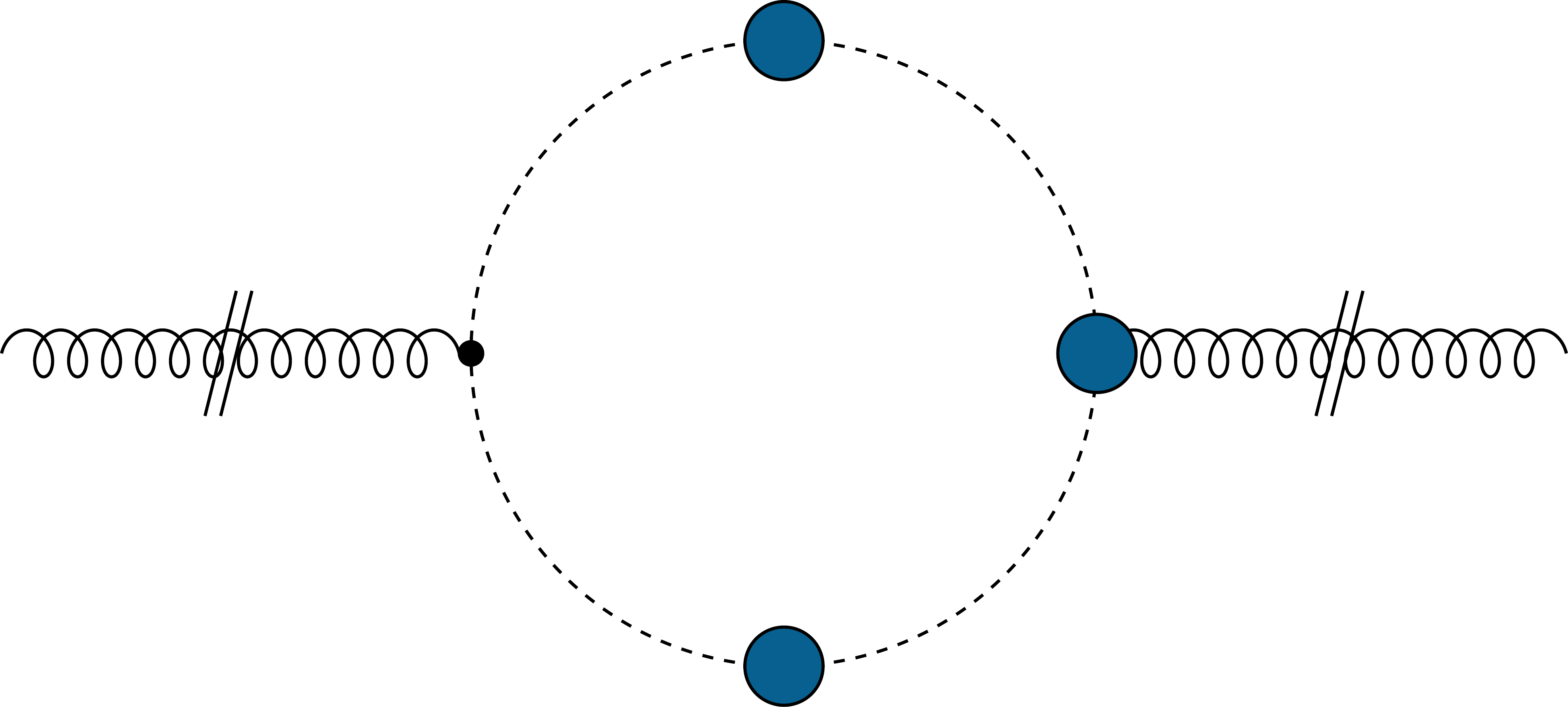}\end{minipage}\,.
\label{curv1}
\end{align}
To the given loop order, the trial measure (\ref{gauss}) becomes exactly Gaussian and 
the variation kernel only enters in the combination $\bar{\omega} = \omega + (1-2 \alpha) \chi$, 
which equals the inverse gluon propagator. (The value of the variational parameters 
$\alpha$ is hence immaterial and, for simplicity, we write 
$\omega(k)$ instead of $\bar{\omega}(k)$ in the following.) The Gaussian is centered at 
$\langle A_\mu \rangle = 0$ as explained above, and we can compute the effective action 
$\Gamma[\mathscr{A} = 0; \omega]$ as a functional of the variational kernel $\omega$. 
Alternatively,
we can view the kernel $\omega$ as the classical value of the inverse gluon propagator
at $\mathscr{A}=0$, and interprete $\Gamma[\mathscr{A} = 0;\omega]$ as its effective action. 
Either way, the \emph{gap equation} $\delta \Gamma / \delta \omega =0$ leads to an integral 
equation for the kernel $\omega$, which can be renormalized and solved together with the 
resolvent identity for the ghost form factor\footnote{By global colour invariance, 
the ghost propagator $\langle G \rangle$ is colour diagonal and the scalar ghost 
form factor can by obtained from the normalized trace in the adjoint representation,
$\mathrm{tr}_A\,\langle G \rangle = \frac{1}{N^2-1}\,\langle G\rangle^{aa}$.}
\begin{align}
\eta(k) \equiv k^2\,\mathrm{tr}_A\,\langle G(k) \rangle
\end{align} 
entering eq.~(\ref{curv1}). For further details, see Refs.~\cite{Quandt:2013wna,Quandt:2016ykm} 
and Fig.~\ref{fig:1}.

These considerations are no longer true at finite temperature. In this case, 
the $O(4)$ spacetime symmetry is manifestly broken since the heat bath singles
out a rest frame. In the imaginary time formalism, the euclidean time interval is 
compactified to the interval $[0,\beta]$ with periodic boundary conditions 
for the gluons and ghosts \cite{Bernard:1974bq}, and we have a residual 
\emph{spatial} $O(3)$ symmetry only. This has various consequences: 
\begin{enumerate}
 \item Integrations over the momentum component $k_0$ are replaced by sums over 
 Matsubara frequencies $k_0 = 2 \pi n / \beta$ with $n \in \mathbb{Z}$,
 \[
\int \frac{d^4 k}{(2\pi)^4}\cdots \quad \Longrightarrow\quad
\int_\beta \dd k\cdots \equiv \beta^{-1}\sum_{n \in \mathbb{Z}} \int
\frac{d^3\mathbf{k}}{(2\pi)^3}\cdots  
 \]
 \item Tensors such as the kernel $\omega_{\mu\nu}^{ab}(k)$ or the 
 curvature $\chi^{ab}_{\mu\nu}(k)$ can now be decomposed into two independent 
 Lorentz structures which are both $4D$ transversal, but $3D$ longitudinal or 
 perpendicular to the heat bath. The gap equation couples the corresponding 
 scalar kernels $\omega_\perp(k_0,|\mathbf{k}|)$ and $\omega_\|(k_0,|\mathbf{k}|)$,
 as well as the ghost form factor $\eta(k_0,|\mathbf{k}|)$ for \emph{all} Matsubara 
 frequencies $k_0 = 2 \pi n/\beta$. The finite temperature system is thus easily 
 several orders of magnitude more expensive than the $T=0$ calculation.
 For practical reasons, we could not include 
 more than about $40$ Matsubara modes in Ref.~\cite{Quandt:2015aaa}, which was limiting 
 the temperature to not much smaller than the critical temperature $T^\ast$. 
 In this temperature range, the resulting propagators 
 agree in all qualitative aspects with the lattice findings.
 In particular,  there is  a slight enhancement of the ghost form factor as compared to 
 $T=0$ and a moderate suppression of the  gluon propagators in the deep infra-red, 
 with higher temperature sensitivity in the components longitudinal to the heat bath.  
 \item The classical field $\mathscr{A}$ which minimizes the effective action need no 
 longer be vanishing. Global colour and spatial $O(3)$ invariance only entail that 
 the minimizing $\mathscr{A}$ can have no spatial component, and the remaining 
 temporal part 
 $\mathscr{A}_0$ should be constant and colour diagonal. In fact, a minimizing 
 classical field  of this type can serve as an alternative order parameter for the 
 deconfinement phase transition, 
 since it obeys the so-called  Polyakov gauge condition, $\partial_0 \mathscr{A}_0 = 0$.
 The most convenient way to introduce a non-vanishing classical field of this type 
 is through the \emph{background field formalism}.
\end{enumerate}

\section{Background field formulation at finite temperature}
\label{sec:background}
Let us elaborate on the last point: in background gauge, we decompose the quantum 
field $A_\mu = \ab_\mu + Q_\mu$, where $\ab_\mu$ is the arbitrary background field, 
and subject the fluctuations to the gauge condition
$[D_\mu(\mathsf{a}), Q_\mu] = 0$, or 
\begin{align}
\hat{\da}_\mu^{ab} Q_\mu^b \equiv 
(\partial_\mu \delta^{ab} - f^{abc} \mathsf{a}_\mu^c) \,Q_\mu^b = 0\,. 
\label{gf1}
\end{align}
Here and in the following, $\da_\mu \equiv D_\mu(\mathsf{a}) = \partial_\mu 
+ \mathsf{a}_\mu$ is the covariant derivative of the background field, and the 
hat over a symbol denotes the adjoint representation.

The effective action for the classical fluctuation field 
$\mathscr{Q}_\mu = \langle Q_\mu \rangle$ depends explicitly on the chosen background field,
$\Gamma_\ab[\mathscr{Q}]$. It is also easy to see that $\Gamma_\ab[\mathscr{Q}]= 
\widetilde{\Gamma}_\ab[\ab + \mathscr{Q}]$, where $\widetilde{\Gamma}_\ab[\mathscr{A}]$
is the usual effective action of the total classical field $\mathscr{A}_\mu$, subject 
to the background gauge condition $\hat{\da}_\mu\,(A_\mu - \ab_\mu) = 0$.
From the considerations in the previous chapter, we know that the minimum of 
$\widetilde{\Gamma}_\ab[\mathscr{A}]$ at any temperature equals the gauge-invariant 
free action, and this minimum occurs at a constant temporal classical field 
$\mathscr{A}_\mu = \delta_{\mu 0}\,\mathscr{A}_0$, provided that the background 
field itself does not break the residual spatial $O(3)$ symmetry. 
To ensure this condition, it is then convenient to choose the background field 
itself in this special form, $\mathsf{a}_\mu = \mathsf{a}_0\,\delta_{\mu 0}$, and 
further set $\mathscr{Q}_\mu = 0$. The minimum of $\widetilde{\Gamma}_\ab[\ab]$ 
for a constant, Abelian background field $\ab_0$ then yields directly the 
gauge-invariant free action. 

\medskip
In Ref.~\cite{Quandt:2016ykm}, we have employed the interpretation 
of $\ab_0$ as an alternative order parameter for the deconfinement phase 
transition \cite{Braun:2007bx,Reinosa:2015gxn}, and evaluated its effective potential
within our approach.
Technically, the transfer of the variational approach from Landau to background gauge is 
rather simple and amounts to the replacement of the partial derivative by its covariant counterpart,
$\partial_\mu \to \hat{\mathsf{d}}_\mu$ in a few strategic places. 
Since the background field $\mathsf{a}_0$ is constant, 
the replacement corresponds to a mere shift in the momentum arguments and we can recycle
the solution of the variational problem in Landau gauge with shifted arguments.
More precisely, $\hat{\da}_\mu$ is a colour matrix in the adjoint representation
and we must first go to a colour base in which $\hat{\da}_\mu$ is diagonal, 
$\hat{\da}_\mu^{ab} = \mathbf{e}^a_\sigma\,(\mathbf{e}^b_\tau)^\ast\,
\delta^{\sigma \tau}\,\mathsf{d}_\mu^\sigma$. This is the so-called \emph{root decomposition}
of the colour algebra, and the momentum shift becomes
\begin{align}
\partial_\mu(p) = i p_\mu \,\,\,\to\,\,\, \da_\mu^\sigma(p) = i (p_\mu - 
\boldsymbol{\sigma} \mathsf{a}_0\,\delta_{\mu 0}) \equiv i p_\mu^\sigma
\label{root}
\end{align}
for every simple root vector $\boldsymbol{\sigma}$. 
We must also replace the factor $(N^2-1)$ from the colour traces in Landau gauge
by a sum over all simple roots.

It should be emphasized that this recipe only holds when using the $T=0$ kernels 
even at finite temperature. The temperature dependent kernels involve the background field 
in other ways than just through the covariant derivative $\da_\mu$, and the 
same also happens if we go beyond two-loop in the effective potential. However, as we have 
seen above, the kernels are only mildly affected by temperature, and it has been further 
argued in Ref.~\cite{Braun:2007bx} that the dominant contributions to the integral 
equations come from momentum and frequency regions where the finite temperature corrections 
to the kernels are negligible. We will thus base our calculations on the $T=0$ 
solutions $\{ \omega(k), \eta(k) \}$ in Landau gauge, cf.~Fig.~\ref{fig:1}. 

\medskip
In background gauge, we therefore replace the measure eq.~(\ref{gauss}) by a similar 
Gaussian \cite{Quandt:2016ykm} centered at $\langle A_\mu \rangle = 
\ab_0\,\delta_{\mu 0}$ and $\langle Q_\mu \rangle = \mathscr{Q}_\mu = 0$
as justified above,
\begin{align}
d\mu &=  \mathcal{N}\cdot \mathcal{D}A\,\mathcal{J}(A)^{1-2\alpha}\,\exp\left[ - \frac{1}{2}\,
\int d(x,y)\,(A - \ab)^a_\mu\,\omega_{\mu\nu}^{ab}(x,y)\,(A - \ab)^b_\nu(y)\right]
\nonumber \\[2mm]
&= \mathcal{N}\cdot \mathcal{D}Q\,\mathcal{J}(\ab + Q)^{1-2\alpha}\,\exp\left[ - \frac{1}{2}\,
\int d(x,y)\,Q^a_\mu(x)\,\omega_{\mu\nu}^{ab}(x,y)\,Q^b_\nu(y)\right]\,.
\label{ansatz}
\end{align}
We can now follow the calculation as in the Landau gauge case: 
in the ghost sector, we have to apply the \emph{rainbow approximation} stating 
that the full ghost-gluon vertex on the r.h.s. of the curvature (\ref{curv1}) is bare.
This simplification is expected to be very robust, since the ghost-gluon vertex is 
known to be non-renormalized 
in Landau gauge due to Taylor's identity \cite{Taylor:1971ff}, and lattice studies also indicate 
that it receives only very mild corrections in the infra-red \cite{Ilgenfritz:2006he,Sternbeck:2006rd}. 
With this approximation, the 
curvature can be expressed solely in terms of the variational kernel $\omega(k)$ and the ghost 
form factor $\eta(k)$, for which a separate resolvent identity exists. The result is the 
free action and the gap equation in background gauge \cite{Quandt:2016ykm}
\begin{align}
F[\ab] &= \frac{1}{2}\,\Big[ \gamma(1,2) + \chi(1,2) + M_0^2(1,2)\Big]\,\omega^{-1}(1,2) + 
\frac{1}{2}\ln \det\omega + \frac{1}{2}\,\ln \det(-\hat{\db}^2)- \ln \mathcal{J}[\ab]\,,
\nonumber \\[2mm]
\omega(1,2) &= \gamma(1,2) + 2 \chi(1,2) + 2 M_0^2(1,2) - \chi_0\,.
\label{gapcool}
\end{align}
To keep the formulas compact, we have used an obvious shorthand notation where a roman 
digit stands for the combination of spactime, Lorentz and adjoint color index, 
$1 \equiv (x, \mu,a),\,\,2 \equiv (y,\nu,b)$ etc., and repeated indices are 
summed or integrated over.\footnote{For instance, the kinetic energy and tadpole 
term from the Yang-Mills action read explicitly,
\begin{align*}
\gamma(1,2) &\equiv \gamma(x,\mu,a\,|\,y,\nu,b) = \Big[- \delta_{\mu\nu}\,\hat{\db}^2_{ab}(x) + 
\hat{\db}_\mu^{ac}(x)\,\hat{\db}_\nu^{cb}\Big]\,\delta(x,y)
\label{gammas}\\[2mm]
M_0^2(1,2) &\equiv M_0^2(x,\mu,a\,|\,y,\nu,b) = 
g^2\,f^{ace}\,f^{ebd}\,\delta_{\mu\nu}\,\big(\omega^{-1}\big)^{cd}_{\alpha\alpha}(x,x)\,
\delta(x,y)\,. 
\end{align*}
}
In momentum space, $M_0^2$ and $\chi_0$ are (quadratically divergent) constants that must 
eventually be removed by renormalization. As layed out in detail in Refs.~\cite{Quandt:2013wna} 
and \cite{Quandt:2016ykm}, we need three counter-terms
\begin{align}
\mathcal{L}_{\rm CT} = \delta Z_A\cdot \frac{1}{4}\big(\partial_\mu A_\nu^a - 
\partial_\nu A_\mu^a\big)^2 + \delta M^2 \cdot \frac{1}{2} \big(A_\mu^a\big)^2 + 
\delta Z_c\cdot \big(\partial_\mu \eta\big)^2\,.
\end{align}
To fix the coefficients, we prescribe the values $Z$ and $M_A$ in the conditions
$\omega(\mu_0) = Z M_A^2$ and $\omega(\mu) = Z \mu^2$ at two different 
scales $\mu_0 \ll \mu$. This removes all quadratic and subleading logarithmic divergences 
from the gap equation.\footnote{Note that the condition for the ``mass counter-term'' is not 
imposed at $\mu_0 = 0$ and $M_A^2$ hence does not have the meaning of a (constituent) mass. 
In fact, the \emph{mass parameter} $M_A^2$ mainly affects the mid-momentum region and also 
appears in the renormalization of the scaling type of solution, where no conventional 
gluon mass emerges.}
In addition, we must also remove the logarithmic divergence in the ghost equation, 
for which we fix of the ghost form factor $\eta(\mu_c) \equiv \eta_0$ at 
$\mu_c \to 0$. The reasoning here is that the present approach allows not only 
for a single solution, but instead for a whole family of scaling and decoupling 
solutions, which differ only in their deep infrared behaviour \cite{Quandt:2013wna}. 
Fixing the ghost form factor at a small scale thus selects a specific type 
of solution and avoids numerical instabilities in the deep infrared.

\begin{figure}[t]
	\centering
	\includegraphics[width=6cm,clip]{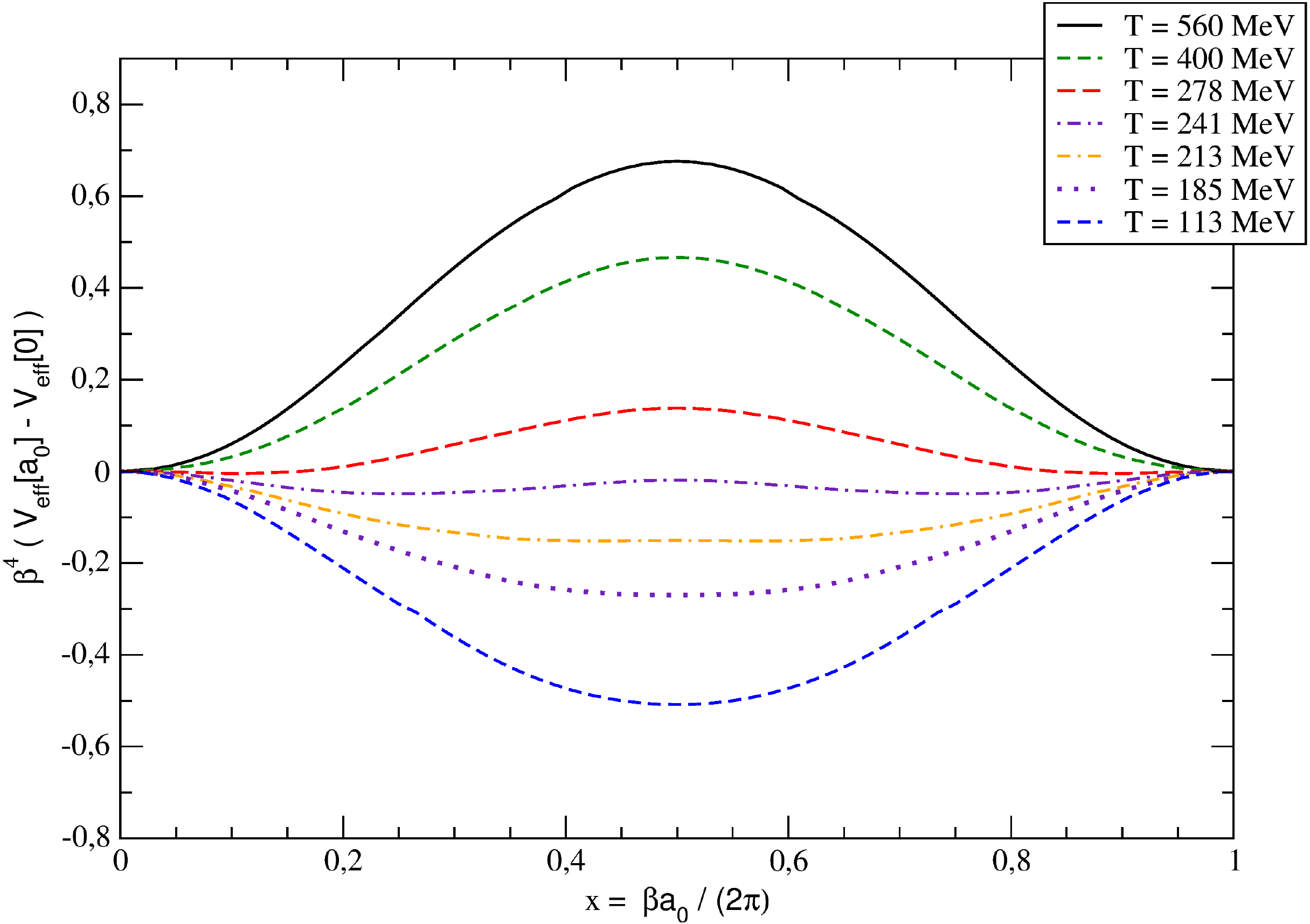}
	\hspace*{1cm}
	\includegraphics[width=6cm,clip]{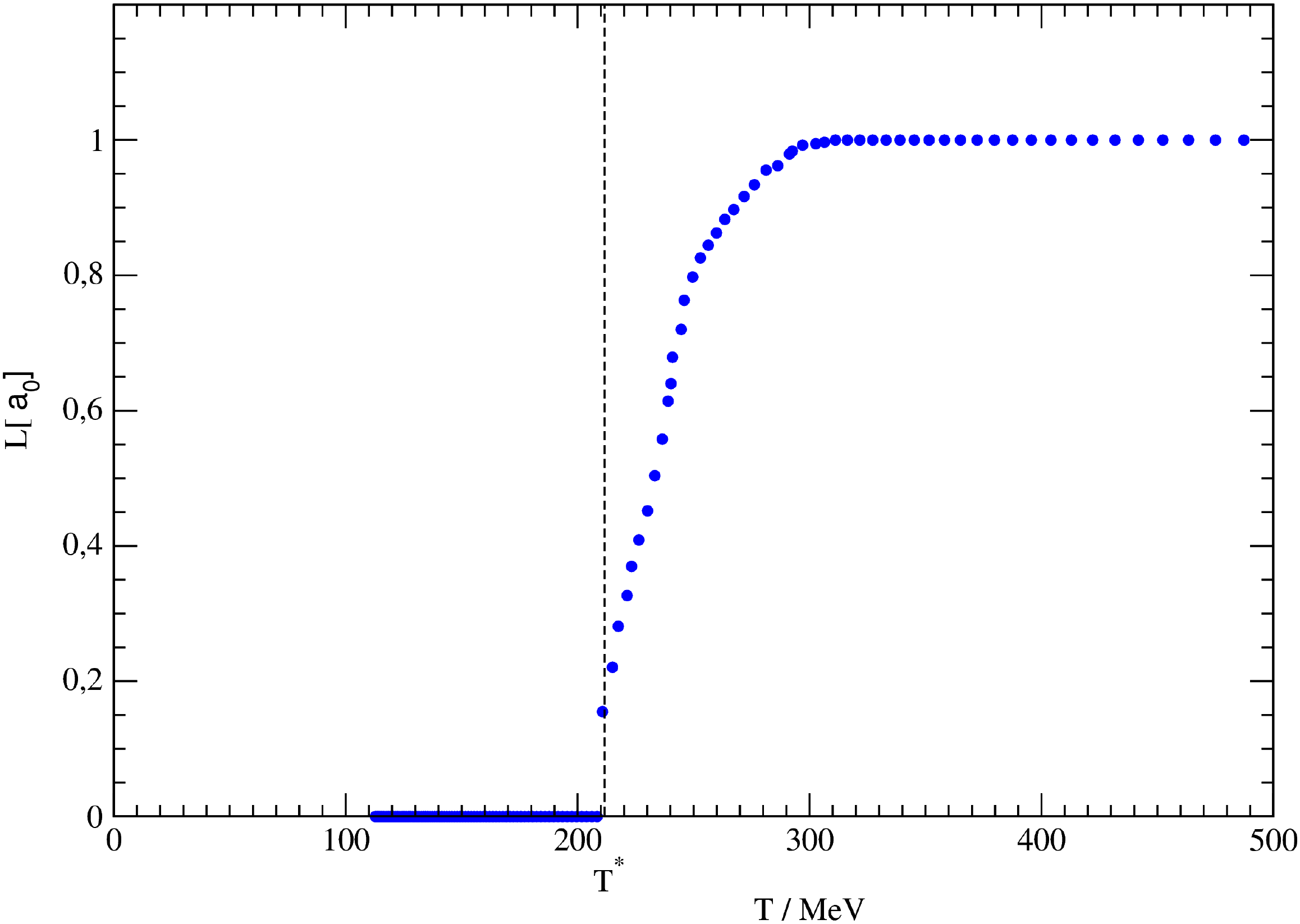}
	\caption{\emph{Left}: The effective potential $V_{\rm eff}(x)$ for the Polyakov loop on the 
		$SU(2)$ Weyl alcove $x \in [0,1]$ defined in eq.~(\ref{su2param}). 
		\emph{Right}: The Polyakov Loop 
		as a function of temperature, extracted from the minimum of $V_{\rm eff}$.}
	\label{fig:3}       
\end{figure}

After renormalization, the mass and wave function counter-terms for the gluon render 
the curvature finite,  $\chi(k) - \chi_0 \to \chi_R(k)$, and the gap equation becomes
\begin{align}
\omega(1,2) = Z\,\gamma(1,2) + 2 \chi_R(1,2)\,. 
\label{gapeq}
\end{align}
If we insert this solution into the free action (with the counter term contribution 
added), we obtain the renormalized effective action of the background field in the 
general form
\begin{align}
 \Gamma[\ab] = - \frac{1}{2}\,\chi_R(1,2) \,\omega^{-1}(1,2) + \frac{1}{2}\ln \det\omega + 
 \frac{1}{2}\,\ln \det(-\hat{\db}^2)- \ln \mathcal{J}[\ab]\,.
 \label{gold}
\end{align}
The kernels $\omega(k)$ and $\chi_R(k)$  are the same functions as computed from the Landau 
gauge system (cf.~Fig.~\ref{fig:1}), but with the momentum argument $k$ shifted by the 
background field in the colour direction of the root vectors, cf.~eq.~(\ref{root}).
For $G=SU(2)$,  there is only a single Cartan generator $T^3 = \sigma^3 / (2i)$ and hence a 
single positive root, and the fundamental domain (Weyl alcove) for the background $\mathsf{a}_0$ 
is conveniently parametrized by the dimensionless quantity 
\begin{align}
x \equiv \beta \mathsf{a}_0^3 / 2\pi \in [0,1]\,.
\label{su2param}
\end{align}
Center symmetry acts as $x \to 1- x$ on these coordinates, and the center symmetric 
point is hence located at $x = 1/2$. At this point, the Polyakov loop $\mathsf{P}$ 
vanishes and we have confinement, while the maximally center breaking 
configurations with $\mathsf{P} = 1$ located at $x=0$ and $x=1$ describe deconfinement.
The deconfinement phase transition thus occurs as a rapid change of the location 
of the minimum of the effective potential $V_{\rm eff}(x)$, from $x=0$ and $x=1$ at $T > T^\ast$ to 
$x=1/2$ at $T \le T^\ast$. This is shown in Fig.~\ref{fig:3}. The transition is 
clearly \emph{second order} for $G=SU(2)$ and the phase transition temperature can be 
translated into absolute units by using the mass scale $M_A^2$ introduced in the $T=0$ propagators. 
This gives a value of $T^\ast \approx 214 \,\mathrm{MeV}$ which is in fair agreement with the lattice 
findings of $T^\ast \approx 300 \,\mathrm{MeV}$ \cite{Lucini:2003zr}, in particular since the 
determination of the scale $M_A^2$ from the fit of the variational solutions to lattice data
has rather large uncertainties. 

Returning to eq.~(\ref{gold}), we can use the curvature representation of the FP determinant
to the given loop order, go to Fourier space and work out the colour traces using the root decomposition, cf.~eq.~(\ref{root}). After introducing spherical coordinates for the 
remaining spatial loop integral, the angles can be integrated out explicitly and we 
obtain
\begin{align}
\beta^4\,f_\beta(x) &= 4 \pi \int_0^\infty dq\,q^2\,
\sum_{n \in \mathbb{Z}}\,\Bigg[ \phi(k_n(x)) + \frac{1}{2}\,\phi(k_n(0)) \Bigg]
\label{basis}
\end{align}
Here, $q \equiv |\vek{k}| \beta / (2\pi)$ is the rescaled dimensionless momentum norm and 
\begin{align}
k_n(x) \equiv \frac{2\pi}{\beta}\sqrt{(n+x)^2 + q^2}\,.
\label{knx}
\end{align}
The integrand in eq.~(\ref{basis}) is given explicitly by
\begin{align}
\phi(k) \equiv 3\,\ln \omega(k) - 3\,\frac{\chi_R(k)}{\omega(k)} - \ln \,k^2\,.
\label{dispersion}
\end{align}
This quantity contains the dispersion relations of all particle fluctuations
in our Gaussian ansatz: we have three non-perturbative transversal 
gluon modes with dispersion $\omega(k)$, the curvature which measures the deviation
of the non-perturbative ghosts from the free ghosts, and the combination of 
these two perturbative ghost modes with the one longitudinal gluon, which 
share the same free dispersion relation $\omega(k) = k^2$.
  
\bigskip  
In principle, the (negative) minimum of eq.~(\ref{basis}) gives the pressure at inverse 
temperature $\beta$, which is our primary goal. However, eq.~(\ref{basis}) is neither in 
a form suitable for numerical evaluation, nor is it finite to start with, even though 
we have added all available counter terms. The reason for this divergence is that 
eq.~(\ref{basis}) still contains the unobservable energy of the vacuum, which must 
be subtracted.\footnote{Formally, this can be associated with a cosmological constant
type of counter term in the original action.} In the present case, this subtraction
becomes particularly transparent if we Poisson resum the Matsubara series in 
eq.~(\ref{basis}),
\begin{align}
f_\beta(x) = \frac{4 \pi}{\beta^4}\,\int_0^\infty dq\,q^2\,\int_{-\infty}^\infty dz\,
\sum_{\nu \in \mathbb{Z}} e^{2\pi i \nu z}\,
\Bigg[\phi\left(\frac{2\pi}{\beta}\,\sqrt{(z+x)^2 + q^2}\right) + \frac{1}{2}\,
\phi\left(\frac{2\pi}{\beta}\,\sqrt{z^2 + q^2}\right)\Bigg]\,.
\label{basis1}
\end{align}
The vacuum energy is precisely the $\nu=0$ contribution to the Poisson sum,
which must hence be omitted to effect the proper subtraction $\delta f$. 
To see this, we can take the $\nu=0$ contribution, shift  $z \to z - x$ in the 
inner integral and change variables $s = z/\beta$ and $p = q / \beta$,
\begin{align}
\delta f = 6\pi\,\int_0^\infty dp\,p^2 \int_{-\infty}^\infty ds\,
\phi\big(2\pi\,\sqrt{s^2 + p^2}\big)\,.
\end{align}
As can be seen, $\delta f$ is a constant contribution to the free action density
independent of the background field $x$ and of temperature $\beta$. This 
(divergent) contribution is present in the free energy density 
$f_\beta(x)$ at any temperature and any background, and can hence be identified
with the constant vacuum energy density or cosmological constant. 

\medskip
After omitting the $\nu=0$ contribution in eq.~(\ref{basis1}), we can further 
evaluate the free energy density using the same analytical techniques that have
have also been used for the effective action of the Polyakov loop in 
Ref.~\cite{Quandt:2016ykm}. For completeness, we have summarized this evaluation 
in appendix \ref{app:A}; the result is the simple expression
\begin{align}
u(x,\beta) \equiv \beta^4\,f_\beta(x) 
= - \frac{2}{\pi^2}\,\sum_{\nu=1}^\infty \frac{\cos(2 \pi \nu x) + \frac{1}{2}}{\nu^4}\,
h(\beta \nu)\,. \label{u}
\end{align}
Here, the function $h(\lambda)$ is the \emph{Hankel transform} of the 
dispersion $\phi(k)$ from eq.~(\ref{dispersion}), 
\begin{align}
h(\lambda) \equiv - \frac{\lambda^3}{4}\,\int_0^\infty dk\,k^2\,J_1(\lambda k)\,\phi(k)\,.
\label{hankel}
\end{align}
As it stands, this transformation is guaranteed to exist for integrands $\phi(k)$ which 
are regular at the origin and vanish sufficiently fast at $k \to \infty$, for 
instance as $\mathcal{O}(k^{-\frac{5}{2}})$. However, we need to extend this domain 
to a larger class of functions $\phi(k)$ with a weaker decay, or even a mild 
(logarithmic) divergence at $k \to \infty$, cf.~eq.~(\ref{dispersion}). To this end, 
we integrate by parts twice and obtain 
\begin{align}
h(\lambda) = \frac{1}{4}\,\int_0^\infty dk\,\Big[2 J_0(\lambda k) - k \lambda J_1(\lambda k)
\Big]\,\big[\phi'(k) + k \phi''(k)\big]\,.
\label{hankel2}
\end{align}
This formula has a much larger domain of convergence and is well suited for numerical evaluation, even when $\phi(k)$ diverges (mildly) at large $k$. The discarding 
of the boundary terms can be justified by distributional arguments, 
cf.~appendix \ref{app:hankel}. As a further test, we have
also considered a free boson of mass $m$, where eq.~(\ref{hankel2}) gives 
the expected result (see also appendix \ref{app:B}),
\begin{align}
h(\lambda) = \frac{1}{2}\,(\lambda m)^2\,K_2(\lambda m)\,.
\label{hmassive}
\end{align}
Eqs.~(\ref{u}) and (\ref{hankel2}) are the main formulas used for 
our numerical study in the next section. Before presenting these results, 
let us first discuss some analytical properties and limits of our approach.

\medskip
In eq.~(\ref{u}), the argument of the Hankel transform is $\lambda = \beta \nu$,
i.e.~the zero temperature limit involves $h_\infty \equiv 
\lim\limits_{\lambda \to \infty} h(\lambda)$. This limit is, however, non-uniform 
with respect to the particle mass. To see this, consider the simple case of a 
free boson of mass $m$ discussed in appendix \ref{app:B}. For any non-vanishing 
mass $m > 0$, we obtain the limit $h_\infty = 0$, while taking $m \to 0$ \emph{first} 
yields $h(\lambda)=1$ for all $\lambda$, and in particular $h_\infty = 1$. 
This means that only massless modes contribute, at very small temperatures, to the rhs 
of eq.~(\ref{u}). Conversely, the high temperature limit $\beta \to 0$ is uniform, 
since $h(0) = 1$ for all free bosonic modes, irrespective of their mass.

\medskip
This observation allows us to easily deduce the high and low temperature limit of the 
rescaled free energy $u(x,\beta)$ from eq.~(\ref{u}): 
As we have seen above, every free massless boson mode is characterized by 
$h(\beta \nu) = 1$ at any temperature, which from eq.~(\ref{u}) gives 
the contribution\footnote{This includes a colour factor $N^2-1 = 3$, since 
all modes are colour fields in the adjoint representation.}
\begin{align}
- \frac{2}{\pi^2}\sum_{\nu=1}^\infty \frac{\cos(2\pi \nu x) + \frac{1}{2}}{\nu^4} = 
\frac{2}{3}\, \pi^2 x^2\,(1-x)^2 - \frac{\pi^2}{30}
\end{align}
to $u(x,\beta)$. Massive free boson modes give the same contribution at high 
energies, while they do not contribute to $u(x,\beta)$ at small temperatures 
$\beta \to \infty$. The proper limit is now a matter of counting degrees of 
freedom: the only massless degrees of freedom at small 
temperatures are the longitudinal gluon and the two  ghost degrees 
of freedom. Together, these represent $(-2) + 1 = -1$ free massless bosonic 
degree of freedom (\emph{ghost dominance}) and we have
\begin{align}
u_\infty(x) \equiv 
\lim_{\beta \to \infty} \beta^4\,f_\beta(x) = - \frac{2}{\pi^2} \sum_{\nu=1}^\infty 
\frac{\cos(2\pi \nu x) + \frac{1}{2}}{\nu^4}\cdot (-1) = -\frac{2}{3}\,\pi^2 x^2(1-x)^2 + 
\frac{\pi^2}{30}\,.
\label{uinf}
\end{align} 
Conversely, all dispersion relations become free  at 
sufficiently high temperatures, and we expect the usual 4 massless gluon 
degrees of freedom, partially cancelled by the two free ghost degrees 
of freedom, for a total of $(+2)$ massless bosonic modes. This gives the limit
\begin{align}
u_0(x) \equiv  \lim_{\beta \to 0} \beta^4\,f_\beta(x) = - \frac{2}{\pi^2} \sum_{\nu=1}^\infty 
\frac{\cos(2\pi \nu x) + \frac{1}{2}}{\nu^4}\cdot (+2) = + \frac{4}{3}\,\pi^2 x^2(1-x)^2 -
\frac{\pi^2}{15}\,.
\label{uzero}
\end{align}
The limits (\ref{uinf}) and (\ref{uzero}) will also be confirmed by the 
numerical study in the next section.

\medskip
The pressure is the negative minimum of the free energy density with respect to the 
background field $x$. For high temperatures, this minimum is reached at $x=0$ 
(\emph{deconfinement}) so that 
\begin{align}
\lim_{T \to \infty} \frac{p(T)}{T^4} = \frac{\pi^2}{15}\,,
\end{align}
which is the correct Stefan-Boltzmann limit for the colour group $G=SU(2)$. 
At very small temperatures, by contrast, $f_\beta(x)$ is minimized at the center 
symmetric background $x=\frac{1}{2}$ (\emph{confinement}), and 
$u_0(\frac{1}{2}) = - \pi^2 / 120$ so that
\begin{align}
\lim_{T \to 0} \frac{p(T)}{T^4} = \frac{\pi^2}{120} \neq 0\,.
\end{align} 

The fact that the pressure (divided by $T^4$) does not vanish at very small 
temperatures is at odds with lattice results \cite{Engels:1988ph,Karsch:2003jg}
and also with general expectations: since the confined phase is characterized 
by a mass gap, all physical excitations above the vacuum (\emph{glue balls})
are much heavier than the critical temperature, so that the free energy and 
the pressure should be exponentially suppressed at $T \to 0$. We will discuss
the origin of this shortcoming (and possible remedies) within our approach 
in section \ref{sec:discuss} below.

\begin{figure}[t]
	\centering
	\includegraphics[width=7.0cm,clip]{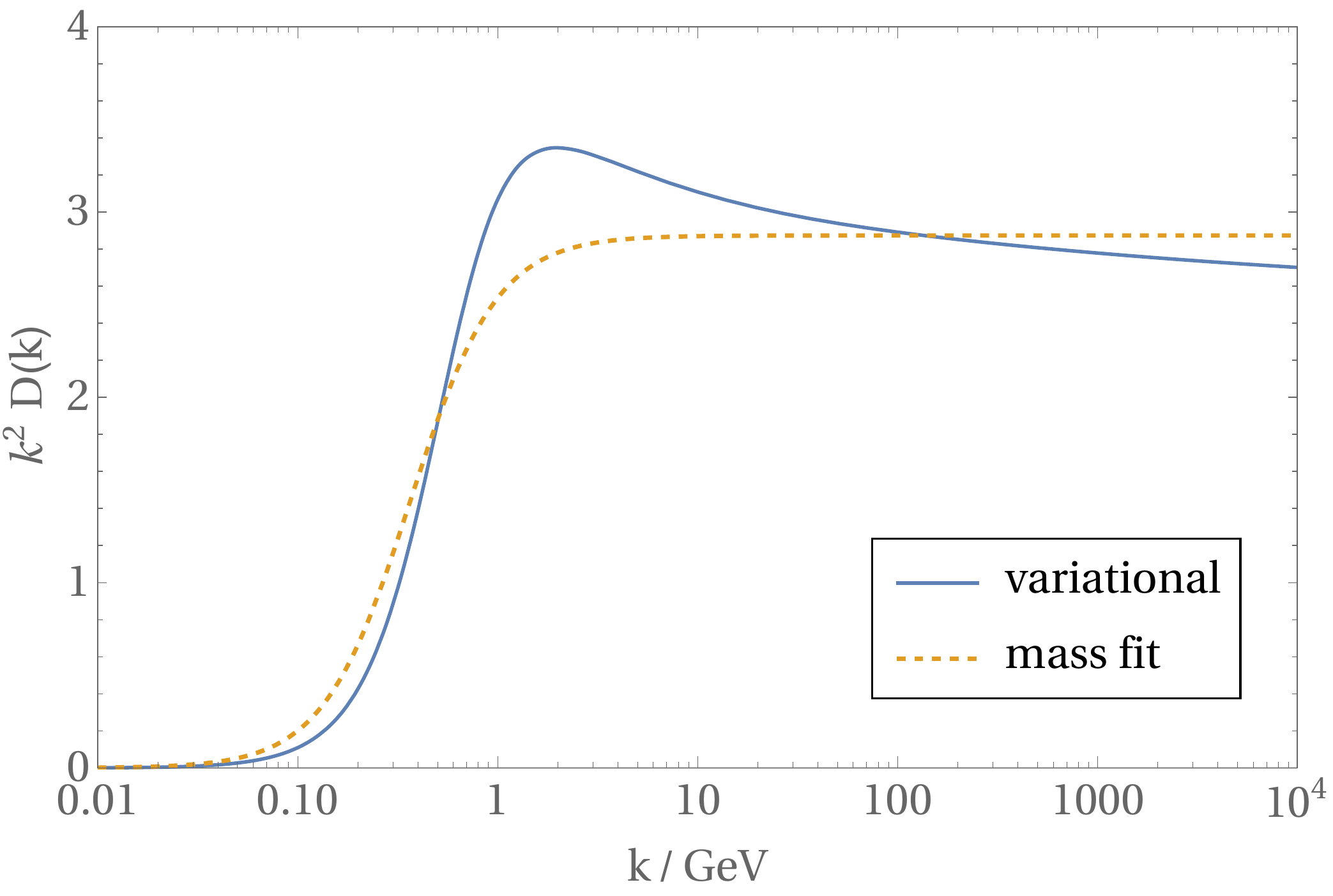}
	\hspace*{0.5cm}
	\includegraphics[width=7.0cm,clip]{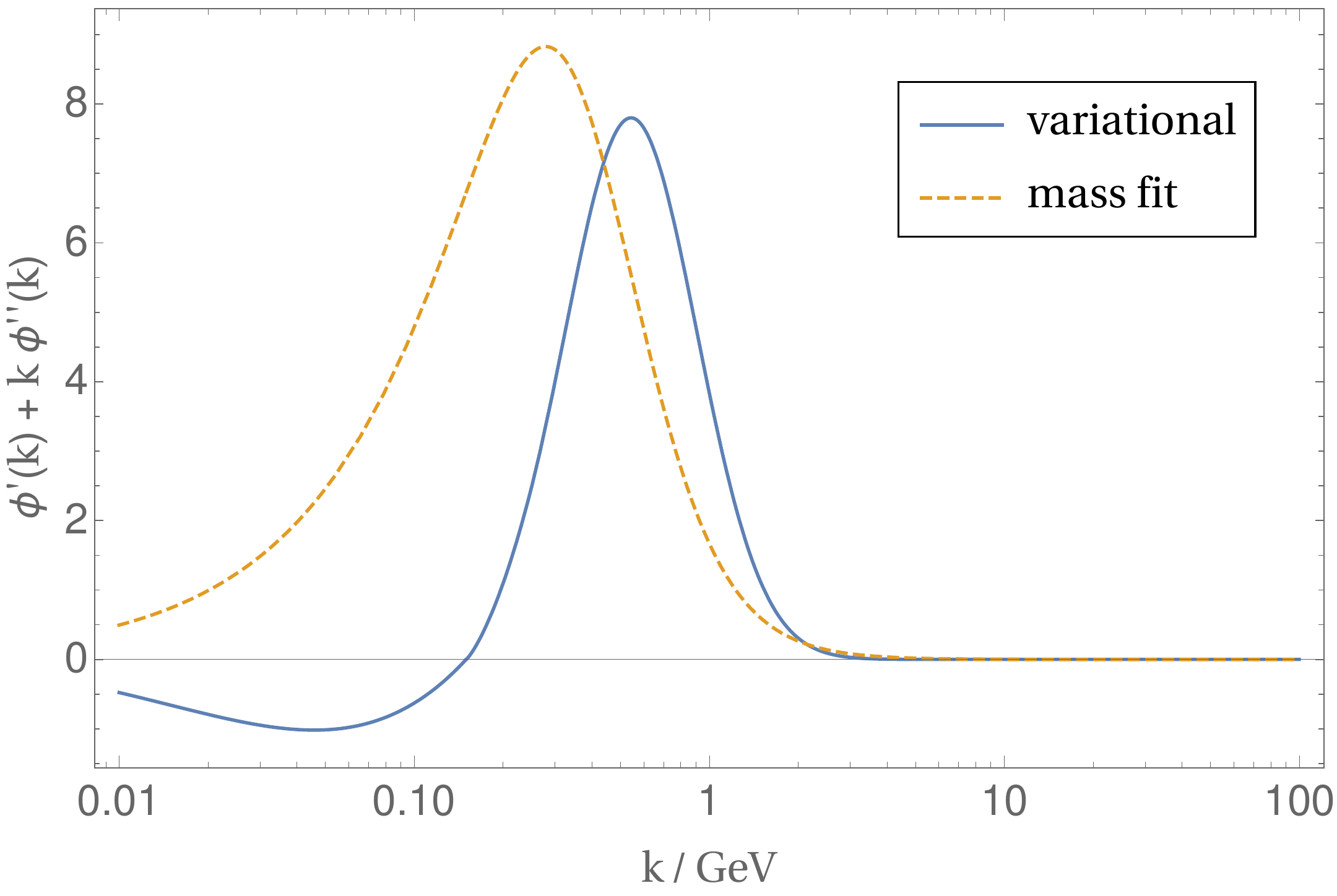}
	\caption{\emph{Left}: The transversal gluon form factor $k^2 / \omega(k)$ of our 
		variational solution, compared to the best fit to the form factor of 
		a massive propagator. \emph{Right}: The integrand $[\phi'(k) + k \phi''(k)]$
		of the Hankel transform eq.~(\ref{hankel2}) \emph{without} the oscillating Bessel
		functions, again compared to the best massive fit to the form factor. }
	\label{fig:4}       
\end{figure}

\section{Results}
\label{sec:results}

\subsection{Free action}
Let us first discuss the calculation of the free energy density as a function 
of the Polyakov loop background field in our variational model.
We take the Poisson resummed formula eq.~(\ref{u}) and compute the Hankel transform 
$h(\lambda)$ from eq.~(\ref{hankel2}), which is also the numerically challenging 
part as the integrand in its definition is strongly oscillating.

\begin{figure}[t]
\centering
\includegraphics[width=7cm,clip]{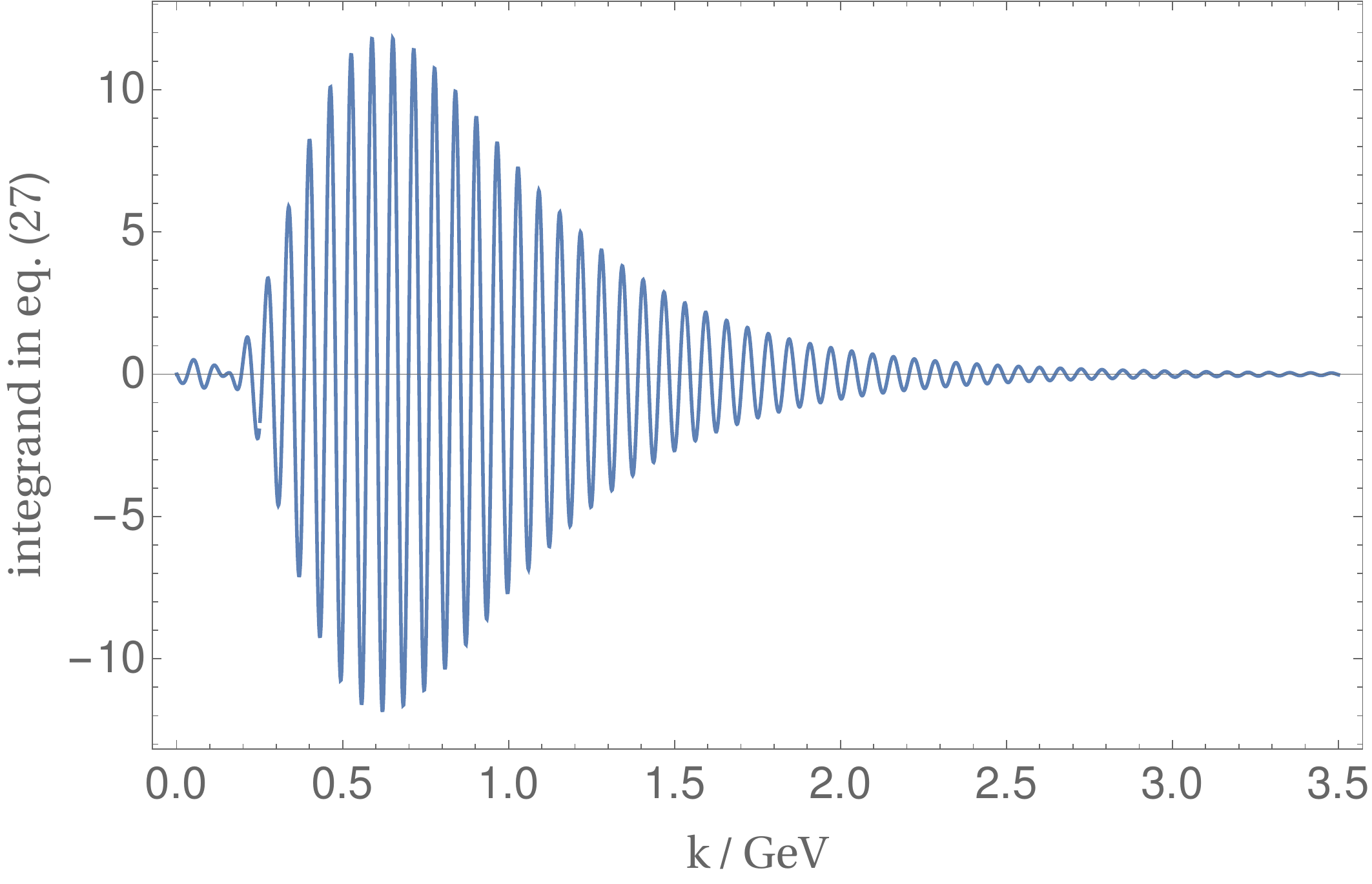}
\hspace*{1cm}
\includegraphics[width=7cm,clip]{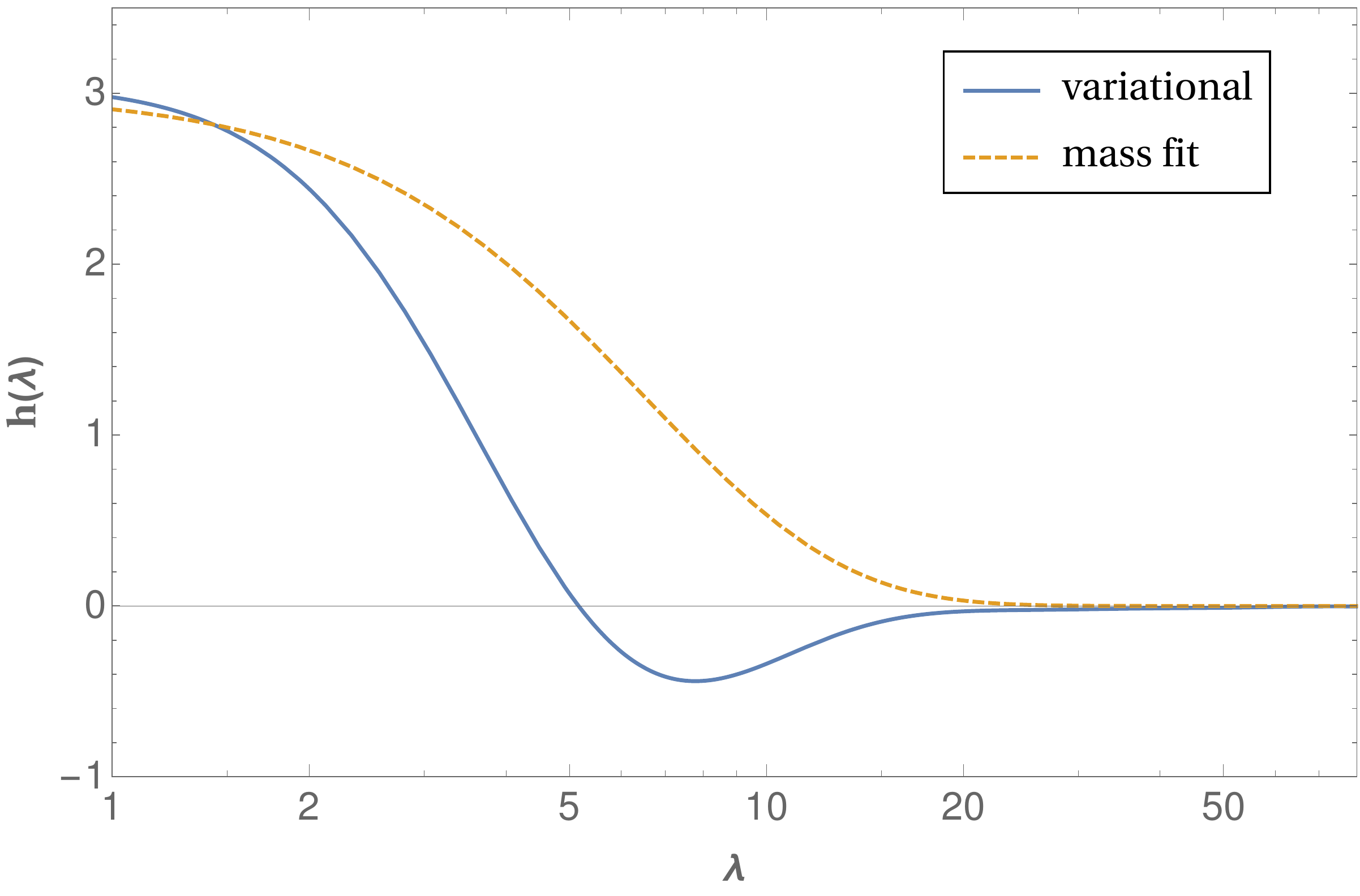}
\caption{\emph{Left}: The full integrand for the Hankel transform in 
eq.~(\ref{hankel2}) for $\lambda = 100$. \emph{Right}: The Hankel transform 
$h(\lambda)$ from eq.~(\ref{hankel2}) for the transversal modes in our 
variational solution, compared to the best massive fit, for which $h(\lambda)$ 
can be calculated analytically.}
\label{fig:5}       
\end{figure}

\medskip
Let us first check whether we can replace the transversal gluon propagator
$\omega(k)^{-1}$ by a massive propagator, since the Hankel transform for the latter 
can be done analytically. In the left panel of Fig.~\ref{fig:4}, we compare 
the form  factor of the transversal propagator, $k^2 / \omega(k)$, for our 
actual numerical solution of the variational problem with the best massive propagator
fit. Clearly, the massive fit underestimates the propagator strength in the mid-momentum 
region, and it also misses the logarithmic decay of the propagator at large 
momenta.
In the right panel of figure \ref{fig:4}, we plot the integrand $[\phi'(k) + k \phi''(k)]$ 
of the Hankel transform eq.~(\ref{hankel2}) \emph{without} the oscillating Bessel 
functions factor. For small momenta, the variational result becomes \emph{negative}, 
a fact that will turn out to be important for the discussion of the pressure 
in the confined phase of our approach. By contrast, the integrand remains positive if 
we replace the propagator by the best fit to a free massive particle. 
\footnote{The right panel of Fig.~\ref{fig:5} only includes the non-perturbative 
contribution from the transversal gluons and the curvature; the latter is taken 
from the gap equation (\ref{gapeq}).}.

\medskip
The full integrand in eq.~(\ref{hankel2}) including the Bessel functions can be wildly 
oscillating, cf.~Fig.~\ref{fig:5}, but the corresponding integral is still convergent
and numerically treatable by standard techniques.\footnote{We change variables 
$k \to q \equiv \lambda k$ to remove the factor $\lambda$ from the argument of the 
Bessel functions, and then split the integral in contributions involving $J_0(q)$ 
and $J_1(q)$. For each integral, we perform a standard quadrature between the (precomputed) 
zeros of the Bessel function and evaluate the resulting alternating sum using 
series accelerators.}
The resulting function $h(\lambda)$ is plotted in the right 
panel of Fig.~\ref{fig:5} for the non-perturbative transversal gluon degrees of 
freedom. (The perturbative ghosts and the longitudinal gluon can be treated 
analytically and merely add constants $\pm 1$ for each mode.)
Clearly, $h(\lambda)$ is bounded, so that the series in eq.~(\ref{u}) is majorized 
by $\sum_\nu \nu^{-4}$ and hence converging very quickly; in practice, we rarely 
need to sum more than $10$ terms to compute $u(x,\beta)$ reliably. 

\begin{figure}[t]
	\centering
	\includegraphics[width=14cm,clip]{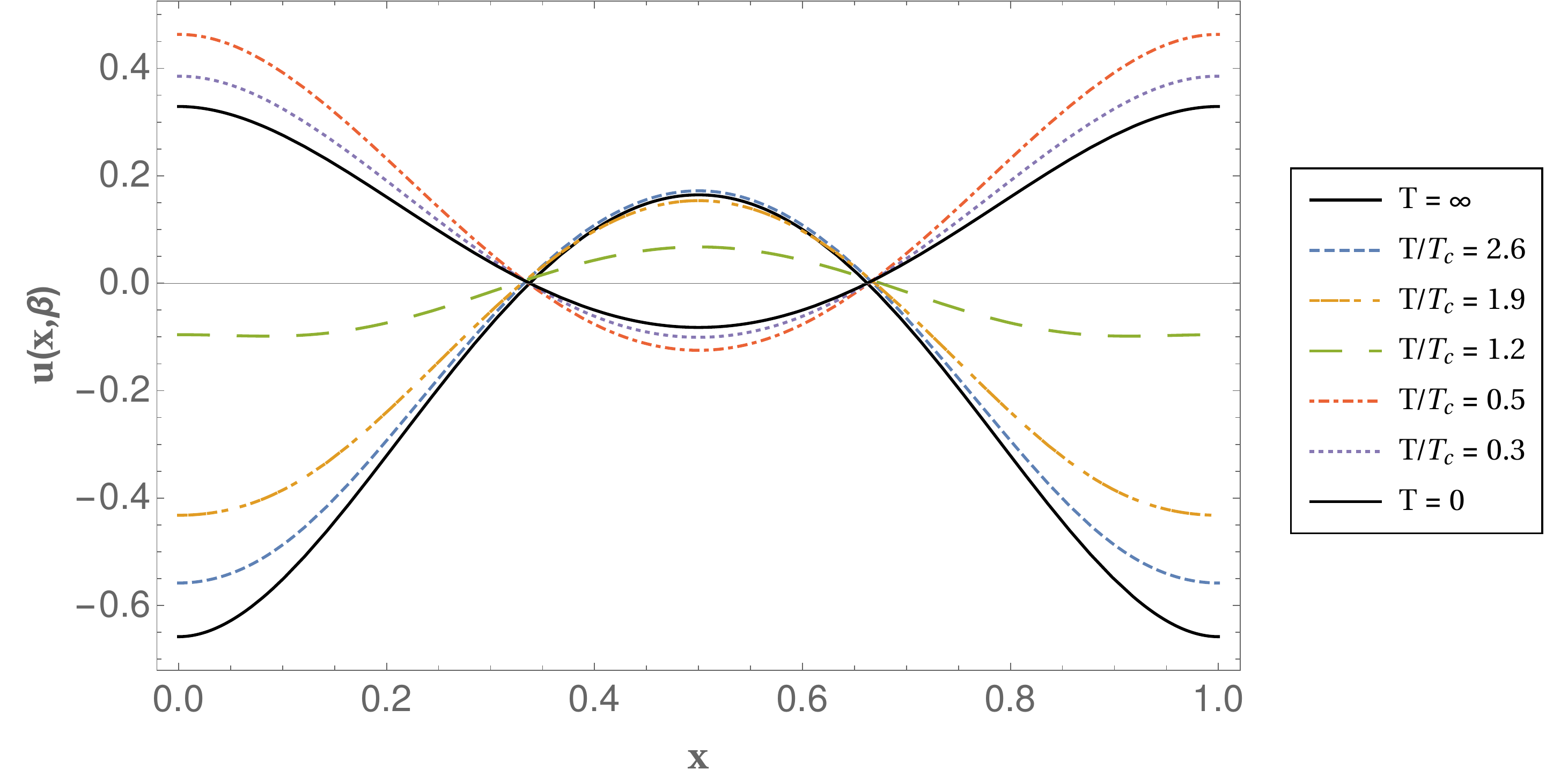}
	\caption{\emph{Left}: The free energy $u(x,\beta) = \beta^4\,f_\beta(x)$ 
		from eq.~(\ref{u}) as a function of the Polyakov loop background for various temperatures. The asymptotic limites at $T\to 0$ and $T \to \infty$ are 
		taken from eqs.~(\ref{uinf}) and (\ref{uzero}), respectively.}
	\label{fig:6}       
\end{figure}

\medskip
Figure \ref{fig:6} shows the free energy $u(x,\beta)$ from eq.~(\ref{u})
as a function of the Polyakov loop background (\ref{su2param}) for various temperatures.
Although the analytical expression eq.~(\ref{uzero}) is clearly the correct
high temperature limit, the approach to this limit is rather slow and still not 
fully reached for temperatures $T \approx 3 T^\ast$. As we lower the 
temperature from the $T\to \infty$ limit, the potential $u(x,\beta)$ quickly moves 
downwards and essentially vanishes at criticality $T=T^\ast$. For even lower temperatures 
(i.e. within the confined phase), $u(x,\beta)$ first reaches the analytical limit 
(\ref{uinf}) at around $T = 0.7\,T^\ast$, before it \emph{overshoots} the limit 
and reaches values well \emph{below} eq.~(\ref{uinf}), before gradually moving 
up again and eventually settling on the limit eq.~(\ref{uinf}) from below. 

The change of shape in $u(x,\beta)$ at $T < T^\ast$ is indispensable to enforce
$x=1/2$ and hence confinement; it is due to the dominating ghost contribution
at small temperatures. Since $u(x,\beta)$ essentially vanishes at criticality, 
the ghost dominance also implies that the free energy must become \emph{negative} 
in the confinement region, which means that there is a positive pressure below 
$T^\ast$, in contrast to lattice calculation and physical expectations. 
In addition, the overshooting of the zero temeprature 
limit $u_\infty(x)$ even induces a shallow maximum of the pressure 
deep within the confining region. This can be traced back to the negative 
section of the Hankel transform $h(\lambda)$ in eq.~(\ref{fig:5}), which in 
turn is related to the negative integrand in Fig.~\ref{fig:4}, and hence to 
the mid- and low-momentum behaviour of the gluon propagator. A free massive 
particle, by contrast, would start from the same high-temperature limit 
$u_0(x)$ for the free energy in Fig.~\ref{fig:6} and also gradually 
move downwards, but eventually will settle on a \emph{flat} profile 
from \emph{above}, without ever turning negative or changing shape;
in particular, it will have the minimum of the free energy at $x=0$ at all 
temperatures and thus show no confinement.  

\subsection{Pressure and thermodynamics}
\begin{figure}[t]
	\centering
	\includegraphics[width=7cm,clip]{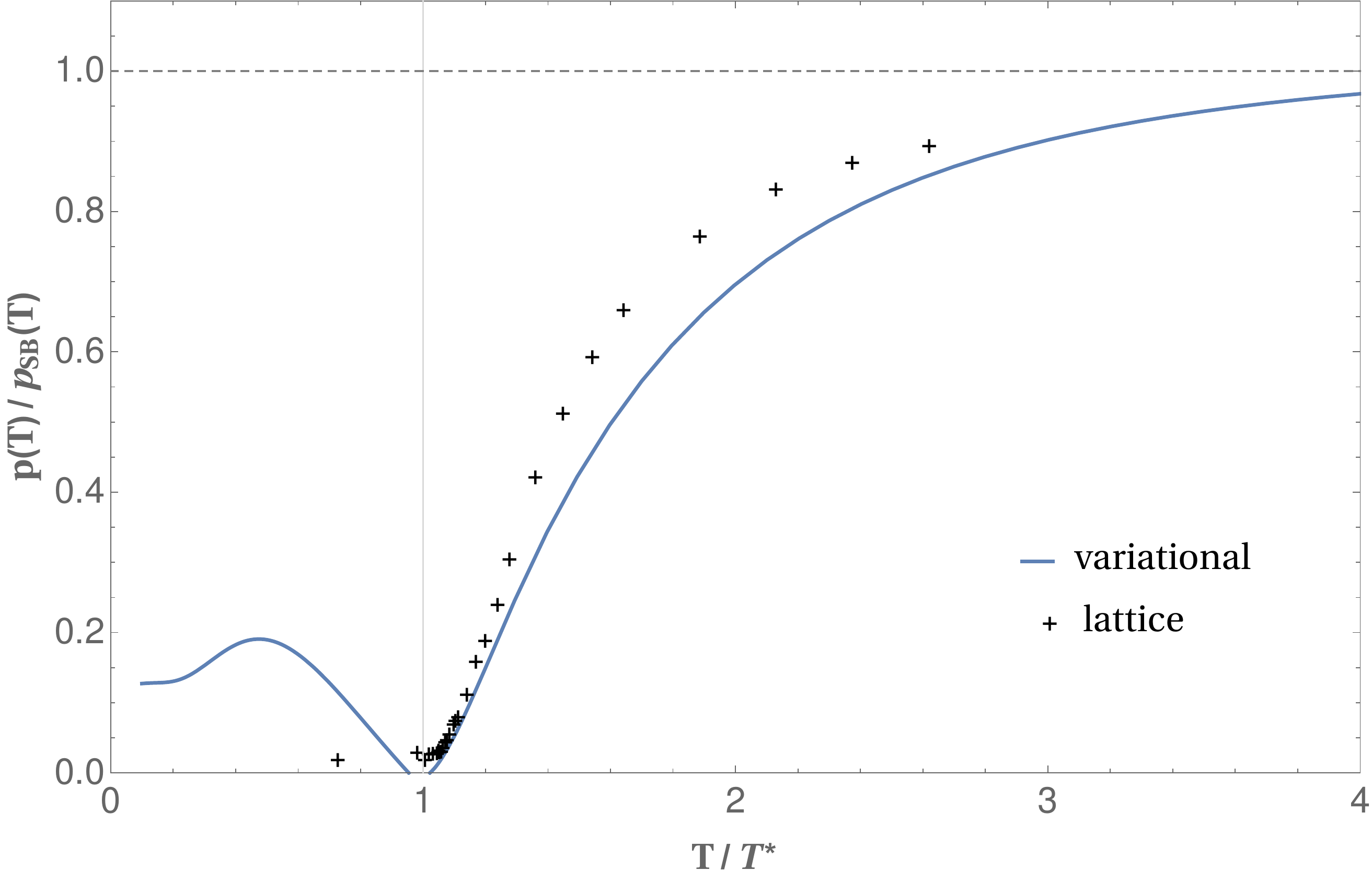}
	\hspace*{1cm}
	\includegraphics[width=7cm,clip]{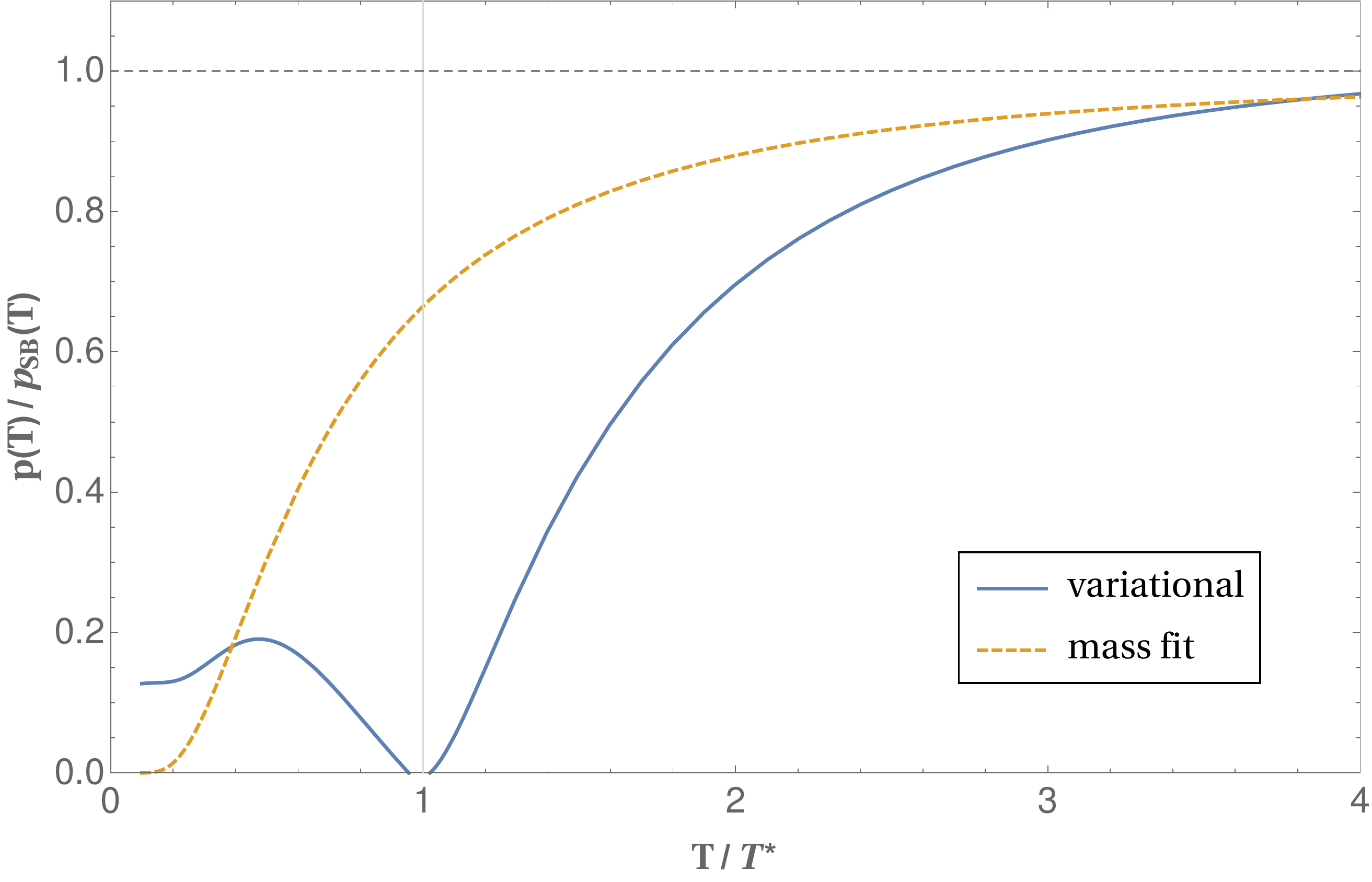}
	\caption{\emph{Left:} The dimensionless pressure $p(T) / T^4$ of the 
		variational solution, normalized to the Stefan-Boltzmann limit.
		In the left panel, our solution is compared to $SU(2)$ lattice data from Ref.~\cite{Engels:1988ph}, while the right panel compares with a model 
		of free massive bosons.}
	\label{fig:7}       
\end{figure}

\begin{figure}[t]
	\centering
	\includegraphics[width=7cm,clip]{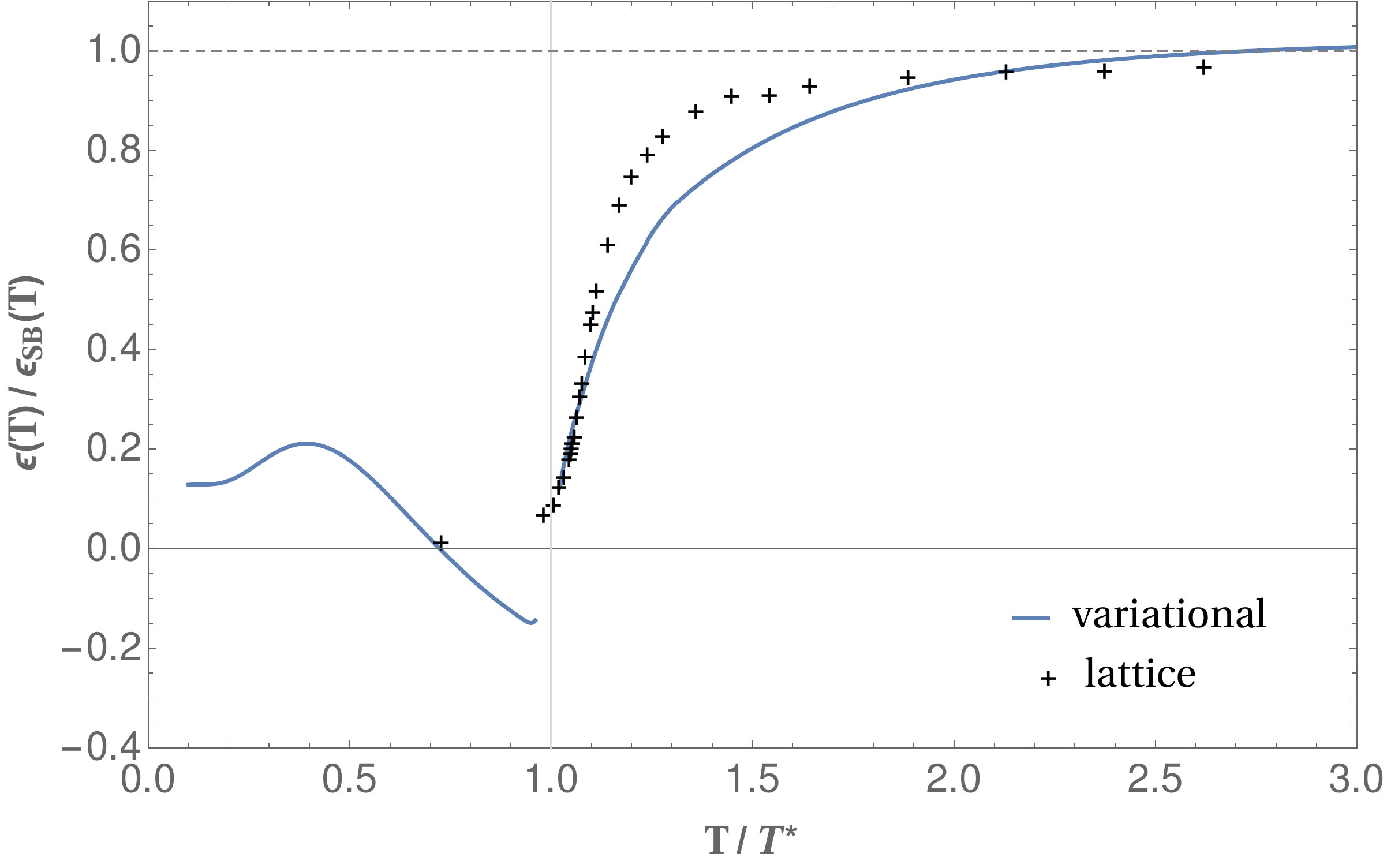}
	\hspace*{1cm}
	\includegraphics[width=7cm,clip]{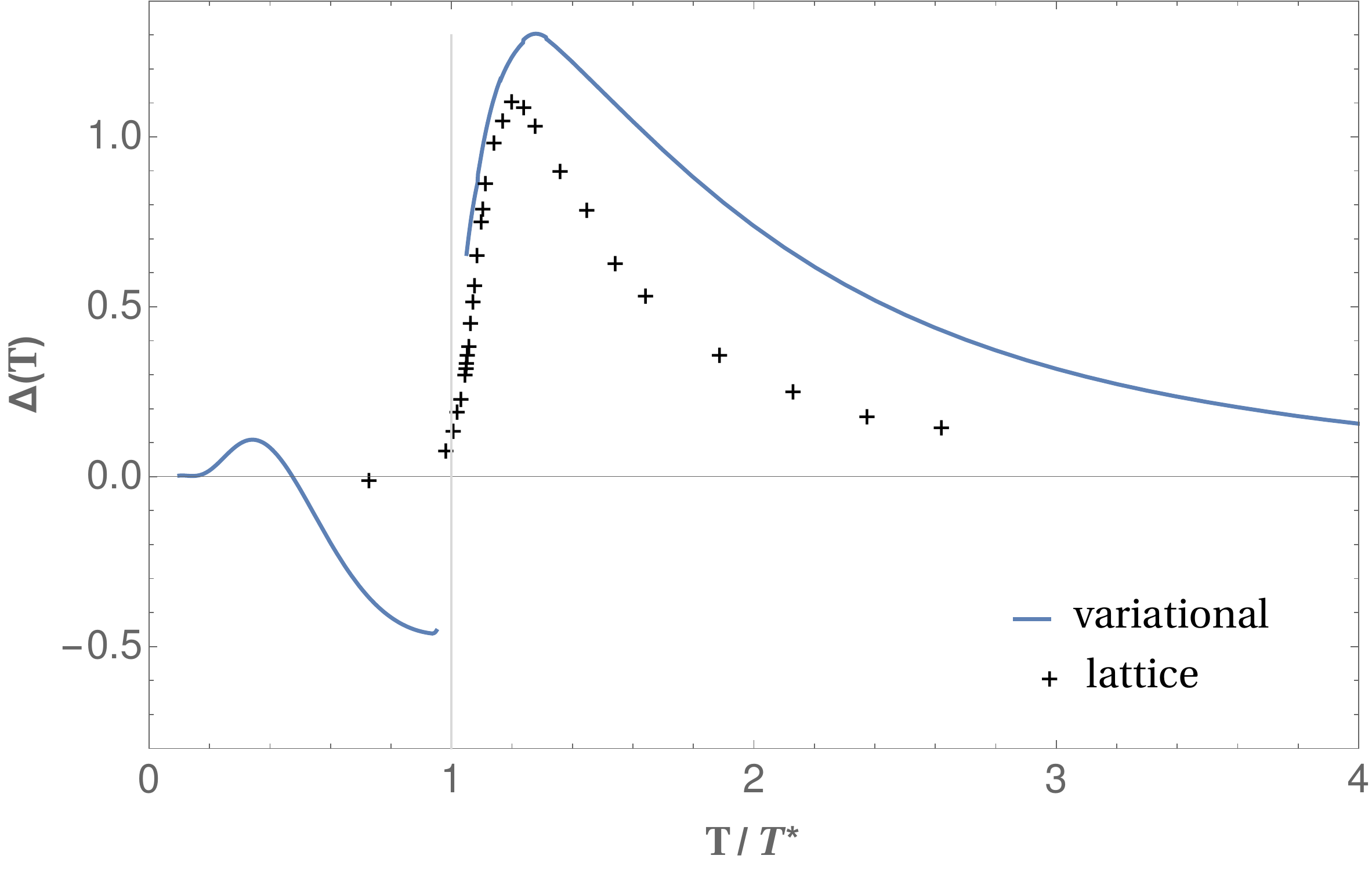}
	\caption{The energy density $\epsilon(T) / T^4$ (\emph{left}) and the 
		interaction strength (\emph{right}) of our variational approach, compared 
		to  $SU(2)$ lattice data from Ref.~\cite{Engels:1988ph}.}
	\label{fig:8}       
\end{figure}

Figure \ref{fig:7} shows the pressure for our variational solution, as computed 
from the negative minimum of the free action density with respect to the Polyakov 
loop background $x$. In the left panel, we compare our solution with the 
SU(2) lattice data from  Ref.~\cite{Engels:1988ph}. We first observe that the 
Stefan-Boltzmann limit at large temperatures is correctly reproduced, as we 
have seen analytically above. 
The agreement with the SU(2) lattice data from Ref.~\cite{Engels:1988ph} is also
fairly good\footnote{For the comparison to the lattice in Fig.~\ref{fig:7} and 
Fig.~\ref{fig:8}, we have fixed the scale of our variational solution by matching
the critical temperature to the lattice. No other parameter was adjusted for all
results in this section.} over the entire deconfinement region 
$T > T^\ast$. At criticality $T=T^\ast$, the pressure vanishes\footnote{The detailed
data exhibits that there might a second zero of the pressure slightly above $T^\ast$,
and hence a very small temperature range in between where the pressure may be
slightly negative. There is a similar finding in Ref.~\cite{Reinosa:2014zta}
(and to a much stronger extent Ref.~\cite{Canfora:2015yia}) where this issue is 
discussed; in our case, the effect is at the brink of our numerical accuracy so 
that we refrain from a detailed discussion.} which is caused by a 
compensation between the transversal gluons and the enhanced ghost degrees of 
freedom. As can be seen in the right panel of Fig.~\ref{fig:7}, a model of 
free massive bosons only would show the correct Stefan-Boltzmann and 
zero-temperature limit, but lacks any sign of a phase transition. 

The situation is more complicated in the confinement region $T < T^\ast$:
Here, the lattice only exhibits the true color-neutral excitations in the 
confined phase, which are glue-balls in the pure Yang-Mills theory, or 
baryons and mesons in full QCD. All these excitations have masses much 
larger than $T^\ast$, so that the free energy (and hence the pressure) 
as well as the energy density become exponentially suppressed in the entire 
confined region $T < T^\ast$. The lattice data in Fig.~\ref{fig:7} confirms 
this expectation, although the data in the confined phase is very scarce. 
By contrast, the pressure in our variational solution \emph{increases} again 
below $T^\ast$, to reach a small maximum around $T \approx 0.7\,T^\ast$, 
before decaying eventually towards the limit 
\begin{align}
\lim_{T \to 0} \frac{p(T)}{p_{\rm SB}(T)} = \frac{1}{8} \neq 0\,.  
\label{baz}
\end{align}
also seen e.g.~in Ref.~\cite{Reinosa:2014zta}.  

The behaviour in the confinement region is problematic since it leads e.g.~to 
negative values of the energy density just below $T^\ast$, and moreover to a 
jump in the energy density not expected for a second order phase transition. 
Similar artifacts can be seen 
in the confined phase of practically all continuum formulations working directly 
with gluon fields, though approaches with a massive behaviour of the transversal 
gluon do not exhibit the small maximum of the pressure around $T \approx 0.7\,T^\ast$
seen in our calculation. In general, the numerical magnitude of the structure in 
the confined region (also quantified by the intercept eq.~()\ref{baz})
is rather small in our case compared to other continuum approaches 
\cite{Canfora:2015yia, Kondo:2015noa}, while our findings do resemble
the results of massive loop calculations \cite{Reinosa:2014zta}.

\medskip
In our approach, we can trace the unusual progression of the pressure in the confined 
region to the following two issues:
\begin{enumerate}
\item The region with a negative value for the Hankel transform
      (cf.~Fig.~\ref{fig:4}) leads to an excessive minimum (overshooting) of 
      the free energy around $T \approx 0.7\,T^\ast$ and hence to the small 
      (unphysical) maximum of the pressure around that temperature. 
\item The non-vanishing value eq.~(\ref{baz}) at zero temperature must be due to 
      the presence of massless modes, since only these modes can contribute 
      to the ratio eq.~(\ref{baz}) at $T \to 0$.
      No such massless excitations are, however, present in the full theory, 
      and on the lattice, below $T^\ast$.
\end{enumerate}

The first issue is, in fact, related to the maximum in the transversal 
gluon form factor at intermediate momenta. This is 
most easily demonstrated by replacing the form factor with a massive 
fit, as indicated in Fig.~\ref{fig:4}. The Hankel transform in
Fig.~\ref{fig:4} then remains positive. If we also keep the longitudinal and 
ghost degrees of freedom as well as the curvature (via the gap equation),
the resulting pressure has a phase transition due to ghost dominance\footnote{The
mass fit is for illustration only and we did not re-adjust its scale 
to actual lattice data. As a consequence, the transition temperature $T^\ast$ 
for the variational solution and mass fit do not match.}, 
but lacks any maximum in the confined region (see Fig.~\ref{fig:9}). 
On the lattice, the maximum in the gluon form factor is also seen, but it 
has no effect on the pressure for $T < T^\ast$ which is thoroughly suppressed 
by the large mass of all physical excitations. To turn this argument 
around, we can say that the presence of massless modes in the confined region 
makes the thermodynamics of our approach more sensitive to details of the 
gluon propagator at intermediate energies than it should.

\subsection{Discussion}
\label{sec:discuss}
As mentioned above, the artifacts and partially unphysical features
in the confined region are predominantly related to the presence of 
massless modes in our approach, even within the confined region. This is, 
to some extent, unavoidable in a continuum calculation, since e.g.~the 
longitudinal gluon will always remain massless and will be present at all 
temperatures. In the exact theory, this gauge mode will receive no 
mass or other radiative corrections beyond one loop, but it will still
be eliminated from the spectrum through cancellation by ghosts via
BRST symmetry. This scenario cannot be fully accommodated in our approach.
We have discussed the lack of BRST symmetry through the Gaussian 
ansatz (not the variation principle itself), and possible remedies or 
improvements in our previous work Refs.~\cite{Quandt:2013wna,Quandt:2015aaa,Quandt:2016ykm}.
Obviously, the Gaussian ansatz is not able to model the real physical 
degrees of freedom (massive glue-balls) in the confined phase. In particular, 
the ghost dominance which is the mechanism for confinement preferred by the 
variational principle, automatically leads to the presence of massless coloured 
degrees of freedom at low temperatures, which in turn induce a non-vanishing 
pressure in the confined phase. A full BRST invariant treatment would 
presumably display a very different confinement mechanism based on colourless, 
massive excitations and BRST invariant states. It is clear that our 
Gaussian ansatz cannot accommodate this scenario and the effect of 
spurious massless gauge modes cannot be fully avoided. These undesirable
consequences of BRST breaking may be less prominent in other observables, 
but the thermodynamics is very sensitive to it, as it predominantly counts
massless degrees of freedom at very small temperatures.

\medskip
Several remedies for this situation come to mind.
The massless modes dominantly affect the Hankel transform limit 
$h_\infty \equiv \lim\limits_{\lambda \to \infty}h(\lambda)$, which should
vanish in the full theory. Since the limit is non-uniform, any 
(infinitesimally small) mass $\mu$ for the longitudinal and ghost modes would enforce the correct 
limit $P(T) / T^4 \to 0$ at $T \to 0$. However, this would only affect the 
temperature region $T < \mu$ and will not solve the issue in the entire 
confined region. What would be required is a rather large mass $\mu \gg T^\ast$
that only exists in the confined region. The variational solution does not 
seem to favour this scenario (and it is forbidden for the longitudinal 
gluon anyway). 

\begin{figure}[t]
	\centering
	\includegraphics[width=7cm,clip]{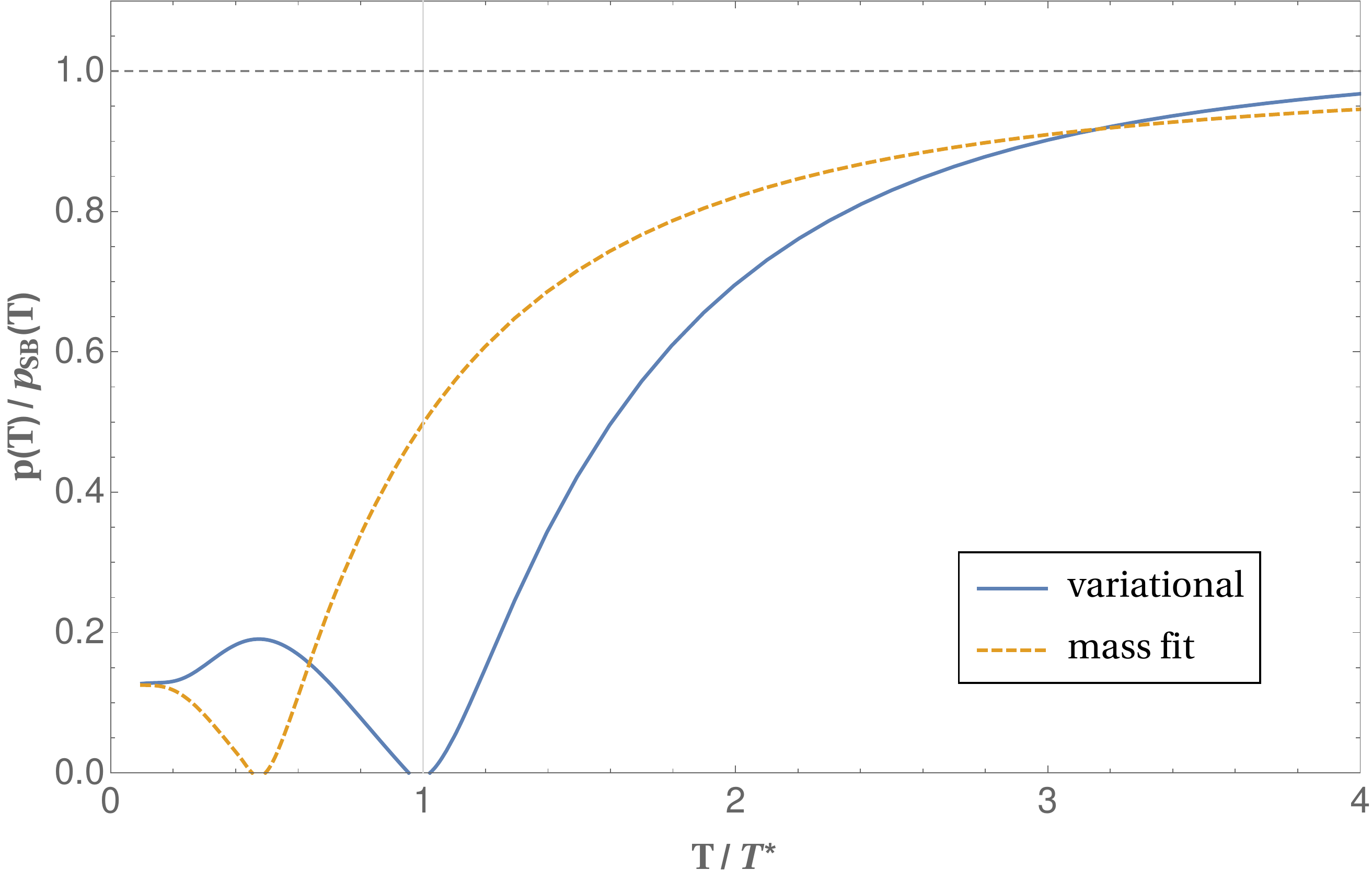}
    \hspace*{1cm}
    \includegraphics[width=7cm,clip]{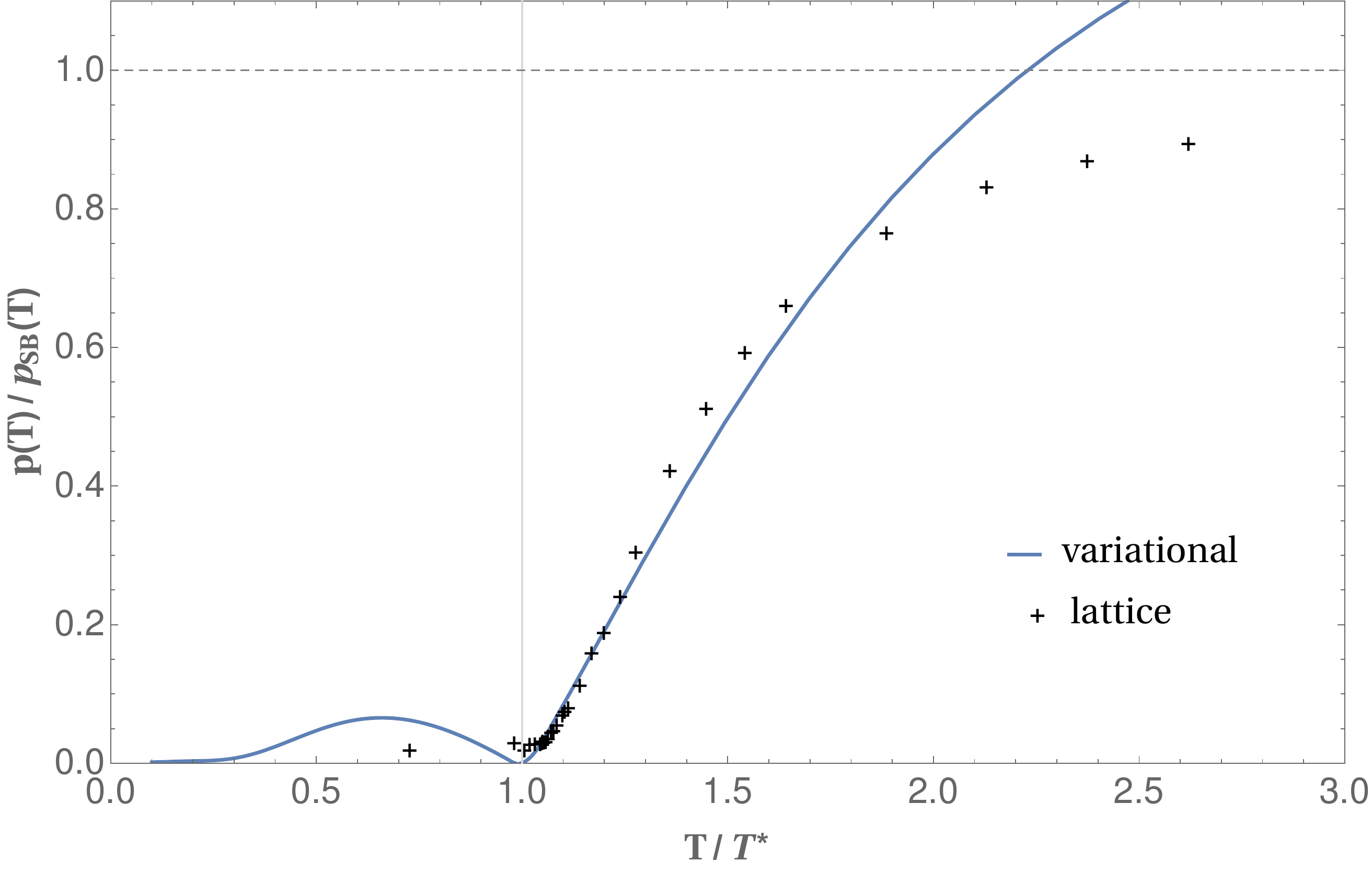}
	\caption{\emph{Left:} The dimensionless pressure $p(T) / T^4$ normalized 
		to the  Stefan-Boltzmann limit, compared to a model in which 
		the transversal gluon form factor has been replaced by a mass fit,
		but longitudinal gluons, ghosts and curvature are fully retained.
	    \emph{Right:} The pressure $P(T) / T^4$ with removal of (spurios)
	    massless modes, compared to lattice data \cite{Engels:1988ph}.}
	\label{fig:9}       
\end{figure}

A second method would be to declare the massless modes as 
unphysical and stipulate that their contribution to $h_\infty$ should 
be removed. This can easily be implemented by subtracting 
$h(\lambda) \to h(\lambda) - h_\infty$, which automatically yields 
$h(\infty) = 0$ and the correct limit at very small temperatures. 
However, this removal is only warranted if the massless modes
have been identified spurious over the entire temperature range,
which is not the case in the deconfined region. 
While the removal of the zero modes gives a very good description of the 
lattice data up to temperatures of about $2 T^\ast$ (see Fig.~\ref{fig:9}),
it will eventually overshoot the Stefan-Boltzmann limit at $T \to \infty$ 
because some of the massless modes in the deconfined region are eliminated 
as well. Clearly, some way of restricting the removal to the confined region
would be necessary, and we see currently no way of implementing this consistently
in the variational ansatz.

Thirdly, we could generally improve on the violation of BRST symmetry by 
extending the variational measure beyond the Gaussian ansatz through either
gluon interaction with non-trivial vertices or even explicit glue ball 
degrees of freedom. The latter could be implemented through additional 
dialton fields e.g.~along the lines of Ref.~\cite{Sasaki:2012bi}.
Non-Gaussian measures, on the other hand, would allow to dress the vertices
in agreement with Slavnov-Taylor identities,  but it could also accommodate 
gluon bound states through non-linear field equations. The techniques for 
dealing with non-Gaussian measures have already been explored in 
Refs.~\cite{Campagnari:2015zsa, Campagnari:2010wc} for the case of the 
Hamiltonian approach. It can straightforwardly adapted to the present case, 
where the gap equations for the vertices provided additional optimizations. 
This third approach is clearly the most physical and its implementation 
will be discussed elsewhere. 

\medskip
In the deconfined region $T \ge T^\ast$, our ansatz displays the right 
degrees of freedom and our variational solution describes the physics 
of the gluon plasma quite accurately. In particular, it exhibits a pressure 
that drops to zero at or near the critical temperature $T^\ast$, and an energy density 
which approaches its Stefan-Boltzmann limit much faster than the pressure 
(compare Fig.~\ref{fig:7} and Fig.~\ref{fig:8}). In addition, it also shows 
a clear maximum of the interaction strength just above the critical 
temperature at around $T \approx 1.25\,T^\ast$, which is only slightly larger 
than the lattice findings of $T = 1.2\,T^\ast$. Both the approach towards 
the Stefan-Boltzmann limit, and the fast decay of the interaction strength at 
higher temperatures are somewhat underestimated in our variational solution,
but the overall agreement with the lattice is still fairly good, given the 
simplicity of our method.

\section{Conclusions}
\label{sec:summary}
In this paper, we have investigated the thermodynamics of $SU(2)$ Yang-Mills
theory in the covariant variational approach. With the properly subtracted 
free energy, we find a pressure for the gluon plasma which yields good 
agreement with the lattice data over the entire deconfinement region, including
the correct Stefan-Boltzmann limit at $T \to \infty$. There is a 
clear phase transition at which the pressure drops to zero, and the interaction
strength shows a pronounced maximum just above $T^\ast$, as is also expected 
from the lattice. In the confined region, the mechanism of ghost dominance and 
spurious massless modes due to the lack of full BRST symmetry make their appearance
and the pressure is not fully suppressed. Instead, it even shows a shallow maximum 
around $0.7\,T^\ast$ before settling to a non-vanishing value for $p(T) / T^4$ 
at $T \to 0$, also in contrast to lattice findings (but in agreement with most 
continuum methods). Such effects are unphysical 
and lead to jumps in the energy density and interaction strength, and even a 
small temperature range of negative energy density. We have attempted to identify
the reason for these problems and discussed possible remedies. To overcome these 
issues, our variational ansatz must be enlarged to include either higher vertices 
and non-linear gluon interactions, or glue-ball degrees of freedom directly. 

A similar extension of the Gaussian ansatz also becomes necessary when 
quark degrees of freedom are introduced and eventually coupled to a 
chemical potential, as the corresponding vertices will be heavily dressed.
The development of Dyson-Schwinger techniques for this formulation in 
analogy to the variational Hamiltonian approach 
\cite{Campagnari:2015zsa, Campagnari:2010wc} 
are currently under investigation.

\begin{acknowledgments}
This work was supported by \emph{Deutsche Forschungsgemeinschaft} under 
contract {DFG~Re856/9--2}.
\end{acknowledgments}


\appendix

\section{Poisson resummation}
\label{app:A}
In this appendix, we derive eq.~(\ref{u}). We start from expression (\ref{basis}) for 
the free action density, which we write
\begin{align}
 \beta^4 f_\beta(x) = 4 \pi \int_0^\infty dq\,q^2 \sum_{n \in \mathbb{Z}}
 \Bigg[\phi\Big(\frac{2\pi}{\beta}\sqrt{(n+x)^2 + q^2}\Big) + 
 \frac{1}{2}\, \phi\Big(\frac{2\pi}{\beta}\sqrt{n^2 + q^2}\Big)\Bigg]\,,
\end{align}
with the function $\phi(k)$ defined in eq.~(\ref{u}). After Poisson resummation,
we obtain
\begin{align}
 \beta^4 f_\beta(x) = 4 \pi \int_0^\infty dq\,q^2 \int_{-\infty}^\infty dz
 \sum_{\nu \in \mathbb{Z}} e^{2 \pi i \nu z}\,
 \Bigg[ \phi\Big(\frac{2\pi}{\beta}\sqrt{(z+x)^2 + q^2}\Big) +
 \frac{1}{2}\, \phi\Big(\frac{2\pi}{\beta}\sqrt{z^2 + q^2}\Big)\Bigg] \,.
\end{align}
As explained in the main text, the $\nu=0$ term describes a 
space-time constant, temperature and background field independent 
contribution $\delta f$, which is part of the free energy density at 
\emph{any} temperature. It can readily be identified as the divergent 
vacuum action density which is to be expected on general grounds and should 
be removed by a cosmological constant type of counter term. After omitting 
this term, we can change variables $s = z + x$ and combine terms with 
$\pm \nu$  to find
\begin{align}
\beta^4 f_\beta(x) 
= 4 \pi \sum_{\nu=1}^\infty &\Bigg[e^{- 2 \pi i \nu x} \int_0^\infty dq \, q^2 
\int_{-\infty}^\infty ds\, \phi\Big(\frac{2\pi}{\beta}\sqrt{s^2 + q^2}\Big)\,e^{2 \pi i \nu s}
\nonumber \\[2mm]
& + e^{2 \pi i \nu x} \int_0^\infty dq \,q^2 
\int_{-\infty}^\infty ds \,\phi\Big(\frac{2\pi}{\beta}\sqrt{s^2 + q^2}\Big)\,e^{-2 \pi i \nu s}
\nonumber \\[2mm]
& + \big(e^{2 \pi i \nu s} + e^{-2 \pi i \nu s}\big)\,\frac{1}{2}\, \int_0^\infty dq\,q^2\int_{-\infty}^\infty ds\,
\phi\Big(\frac{2\pi}{\beta}\sqrt{s^2 + q^2}\Big)
\Bigg]\,.
\end{align}
Changing variables $s \to (-s)$ in the second and third term gives
\begin{align}
\beta^4 f_\beta(x) =  8 \pi \sum_{\nu=1}^\infty\Big[\cos(2\pi \nu x) + \frac{1}{2}\Big]
\int_0^\infty dq q^2 \int_{-\infty}^\infty ds\, \phi\Big(\frac{2\pi}{\beta}
\sqrt{s^2 + q^2}\Big)\,e^{2 \pi i \nu s}\,.
\end{align}
The integrand is even in $q$ and we can extend the $q$-integral to all of $\mathbb{R}$. Next 
we introduce polar coordinates $(r,\varphi)$ in the $(q,s)$-plane and find
\begin{align}
\beta^4 f_\beta(x)  &=  4 \pi \sum_{\nu=1}^\infty\Big[\cos(2\pi \nu x) + \frac{1}{2}\Big]
\int_0^\infty dr\, r \int_0^{2\pi} d\varphi\,r^2 \sin^2 \varphi \cdot 
\phi\big(2 \pi r / \beta\big)\,e^{2 \pi i r \nu \cos\varphi}
\nonumber \\[2mm]
&=4 \pi \sum_{\nu=1}^\infty\Big[\cos(2\pi \nu x) + \frac{1}{2}\Big]
\int_0^\infty dr \,r^3\,\phi\big(2 \pi r / \beta\big)\,
\int_0^{2\pi} d\varphi\,\sin^2\varphi\cdot e^{2 \pi i r \nu \cos\varphi}
\nonumber \\[2mm]
&= 4\pi \sum_{\nu=1}^\infty\Big[\cos(2\pi \nu x) + \frac{1}{2}\Big]
\int_0^\infty dr\, r^2\,\phi\big(2 \pi r / \beta\big)\,
\frac{J_1(2 \pi \nu r)}{\nu}\,.
\end{align}
A final change of variables $\tau \equiv 2 \pi \nu r$ yields
\begin{align}
\beta^4 f_\beta(x) &= \frac{1}{2\pi^2}\sum_{\nu=1}^\infty
\frac{\cos(2\pi \nu x) + \frac{1}{2}}{\nu^4}\int_0^\infty d\tau\,
\tau^2 \,J_1(\tau)\,\phi(\tau / (\nu \,\beta))\,.
\end{align}
If we abbreviate the Hankel transform by
\begin{align}
h(\lambda) =  - \frac{1}{4}\,\int_0^\infty d\tau\,\tau^2\,J_1(\tau)\,\phi(\tau/ \lambda)
\end{align}
we arrive eventually at eq.~(\ref{u}) from the main text.

\section{Hankel transformation}
\label{app:hankel}
We want to compute the Hankel transformation eq.~(\ref{hankel}) 
in the main text, 
\begin{align}
h(\lambda) = - \frac{\lambda^3}{4} \int_0^\infty dk\,k^2\,J_1(\lambda k)\,\phi(k)\,, 
\label{tea}
\end{align}
where we first assume that the integrand $\phi(k)$ is regular at the origin and 
vanishes fast enough at large $k$ so that the integral converges.
In this case, we can integrate by parts twice to obtain
\begin{align}
h(\lambda) = \frac{1}{4}\,\int_0^\infty dk\,\left(2 J_0(\lambda k) + \lambda k \,J_1(\lambda k)
\right)\cdot \left(k \phi''(k) + \phi'(k)\right)\,,
\label{pot}
\end{align}
which is eq.~(\ref{hankel2}) from the main text. This formula has the advantage, 
that it converges for functions $\phi(k)$ which have a much weaker decay at large 
$k$, or even a mild (logarithmic) increase\footnote{In 
this case, eq.~(\ref{pot}) leads to convergent integrals of the  
Hankel-Nicholson type, cf.~appendix \ref{app:B}.} at $k \to \infty$.
 
We can view eq.~(\ref{pot}) as an analytical continuation of the initial 
formula (\ref{tea}) which can be justified by density arguments in many cases. 
More explicitly, we can regularize the original integral (\ref{tea}) in 
a distributional sense by replacing $\phi(k) \to \phi(k) \cdot e^{-\mu k}$ 
with $\mu > 0$, and consider
\begin{align}
h(\lambda) = \lim_{\mu \to +0}
 \frac{\lambda^3}{4} \int_0^\infty dk\,k^2\,J_1(\lambda k)\,\phi(k)\cdot e^{-\mu k}\,.
 \label{hankelreg}
\end{align}
Integrating by parts twice yields for the integral without the limit
\begin{align}
-& \Big[ \frac{(\lambda k)^2}{4} \,J_2(\lambda k)\,\phi(k) + 
\frac{k}{4}\,[\phi'(k)- \mu \phi(k)]\cdot[2 J_0(\lambda k) + \lambda k\,J_1(\lambda k)]
\Big]\cdot e^{-\mu k}\Bigg\vert_0^\infty
\nonumber \\[2mm]
+& \frac{1}{4}\int_0^\infty dk\,[2 J_0(\lambda k) + \lambda k\,J_1(\lambda k)]\cdot 
[k \phi''(k) + (1- 2 \mu k) \phi'(k) + \mu (\mu k -1)\,\phi(k)]\cdot e^{-\mu k}\,.
\label{hankelfoo}
\end{align}
If $\phi(k)$ is regular at the origin, the boundary contribution from $k=0$ vanishes;
likewise, the boundary contribution from $k=\infty$ vanishes due to the common 
factor $e^{-\mu k}$, provided that $\phi(k)$ only has a mild divergence
at $k \to \infty$. In this case, 
\begin{align*}
(\ref{hankelreg}) = \lim_{\mu \to +0} \frac{1}{4} \int_0^\infty dk\,[2 J_0(\lambda k) + \lambda k\,J_1(\lambda k)]\cdot [k \phi''(k) + (1- 2 \mu k) \phi'(k) + \mu (\mu k -1)\,\phi(k)]\cdot e^{-\mu k}
\end{align*}
and the limit can be done under the integral, which leads to eq.~(\ref{pot}). 

We note in passing that the distributional limit is important: if we put $\mu=0$ 
in eq.~(\ref{hankelreg}) and instead regularize the integral by a cutoff $\Lambda$, 
then the boundary term in eq.~(\ref{hankelfoo}) will not vanish
unless the the cutoff is taken to infinity as the special sequence
$\Lambda_n = u_n / \lambda$, where $u_n\approx\pi(\frac{5}{4} + n)$ is the nth zero 
of the Bessel function $J_2(x)$. For this sequence of cutoffs $\Lambda_n \to \infty$,
the first boundary term in eq.~(\ref{hankelfoo}) vanishes, while the second one 
becomes independent of $\lambda$. The result would then again be eq.~(\ref{pot}),
however with an additional constant $h_0$ on the rhs, which only vanishes if 
$\phi'(k) \sim k^{-(1+\epsilon)}$ at large $k$. The distributional argument has the 
merit of extending eq.~(\ref{pot}) to a wider class of functions with 
stronger divergence at $k \to \infty$.

\section{Massive free bosons}
\label{app:B}
To check the results from appendix \ref{app:hankel}, let us study a single scalar 
boson of mass $m$ with dispersion relation $\omega(k) = k^2 + m^2$, i.e.~we consider
\[
 \phi(k) = \ln \frac{\omega(k)}{\mu^2} = \ln \frac{k^2 + m^2}{\mu^2}\,,
\]
where the scale $\mu$ is arbitrary and not to be confused with the regulator in 
appendix \ref{app:hankel}. The model function $\phi(k)$ has only a mild 
(logarithmic) divergence at $k \to \infty$ and eq.~(\ref{pot}) leads to a
Hankel-Nicholson type of integral,
\begin{align}
h(\lambda) &= \frac{1}{4}\,\int_0^\infty dk\,
\Big[2 J_0(\lambda k) + \lambda k \,J_1(\lambda k)\Big] \cdot 
\frac{4 k m^2}{(k^2 + m^2)^2}
\nonumber \\[2mm]
&= (\lambda m)^2 \,\int_0^\infty dq\,
\big(2 J_0(q) + q J_1(q))\,\frac{q}{(q^2 + (\lambda m)^2)^2}
\nonumber \\[2mm]
&= \frac{1}{2}\,(\lambda m)^2\,K_2(\lambda m)\,.
\label{foxy}
\end{align}
Here, $K_2$ is a modified Bessel function and the result is indeed independent
of the scale $\mu$. In the massless limit $m \to 0$, we recover the simple 
result $h(\lambda) = 1$. 
\bibliographystyle{apsrev4-1}
\bibliography{thermobib}
\end{document}